\documentclass[10pt,twocolumn]{article}

\usepackage{newtxtext,newtxmath}

\usepackage{graphicx}

\usepackage[letterpaper,margin=0.5in,columnsep=0.2in]{geometry}

\date{}

\renewcommand\refname{References and Notes}

\makeatletter
\renewcommand{\fnum@figure}{\textbf{Fig. \thefigure}}
\renewcommand{\fnum@table}{\textbf{Table \thetable}}
\makeatother

\usepackage{scicite}
\usepackage{pdfrender}
\usepackage{url}
\usepackage{caption}
\DeclareCaptionFont{phv}{\fontfamily{phv}\selectfont}
\usepackage{fancyhdr}
\usepackage{lastpage}

\def\scititle{
	Mode-selective acousto-electric modulation of phonons in a silicon photonic platform
}
\title{\bfseries \boldmath \scititle}

\usepackage{xcolor}
\definecolor{saRed}{RGB}{170,30,45} 
\usepackage{dblfloatfix}
\begin{document}

\twocolumn[

\begin{center}
{\LARGE\bfseries \scititle \par}
\vspace{0.8em}
{\normalsize\bfseries
Ruoyu Yuan$^{1}$, Yishu Zhou$^{2}$, Matthew J. Storey$^{3}$, Ryan O. Behunin$^{4,5}$,
Haotian Cheng$^{2}$, Betul Sen$^{2}$, Andrew L. Starbuck$^{3}$, Douglas C. Trotter$^{3}$,
Andrew L. Leenheer$^{3}$, Matt Eichenfield$^{3,6,7}$, Nils T. Otterstrom$^{3}$,
Peter T. Rakich$^{2*}$
\par}
\end{center}

\vspace{0.6em}

\begin{center}
\parbox{0.95\textwidth}{
\noindent
\textbf{Acousto-electric (AE) interactions enable electrical control of acoustic propagation through piezoelectric media. Bringing AE control onto integrated photonic platforms provides a powerful on-chip control mechanism to reconfigure both the acoustic delay-line response and the effective photon–phonon interaction by electrically tuning the phonon propagation. Here we report a mode-selective AE modulation in a scalable aluminum nitride on silicon-on-insulator (AlN-on-SOI) platform. An applied DC field is seen to modulate the phonon dissipation via changes in carrier concentration in a mode-selective fashion, producing up to 20 dB forward transmission suppression of non-fundamental modes within an acoustic delay line while maintaining the fundamental Rayleigh mode, a phenomenon not captured by conventional AE treatments. To probe these dynamics, we integrate an optical waveguide along the acoustic delay line, providing a broadband, non-destructive interface between the acoustic and optical domains. Leveraging both this optical interface and the underlying mode selectivity, we demonstrate proof-of-concept multi-domain transduction and Rayleigh-mode filtering on this hybrid platform, outlining a scalable path toward electrically reconfigurable mode engineering and microwave–photonic functionality.}
}
\end{center}
\vspace{1em}
]
\renewcommand{\footnoterule}{}
\begingroup
\renewcommand\thefootnote{}
\footnotetext{
\noindent\hspace*{-1.7em}\rule{\columnwidth}{0.4pt}\par
\hspace{-1em}\parbox{\columnwidth}{\fontfamily{phv}\footnotesize
$^{1}$Department of Electrical Engineering, Yale University, New Haven \& 06511, USA.\\
$^{2}$Department of Applied Physics, Yale University, New Haven \& 06511, USA.\\
$^{3}$Sandia National Laboratories, Albuquerque, New Mexico, USA.\\
$^{4}$Department of Physics, Northern Arizona University, Flagstaff, Arizona \& 86011, USA.\\
$^{5}$Center for Materials Interfaces in Research and Applications, Northern Arizona University, Flagstaff, Arizona \& 86011, USA.\\
$^{6}$James C. Wyant College of Optical Sciences, University of Arizona, Tucson, Arizona \& 85716, USA.\\
$^{7}$Electrical, Computer and Energy Engineering, University of Colorado Boulder, Boulder, Colorado \& 80309, USA.\\
$^{*}$Corresponding author. Email: peter.rakich@yale.edu
}
}
\endgroup

\begin{figure*}[!b] 
	\centering
	\includegraphics[width=0.85\textwidth]{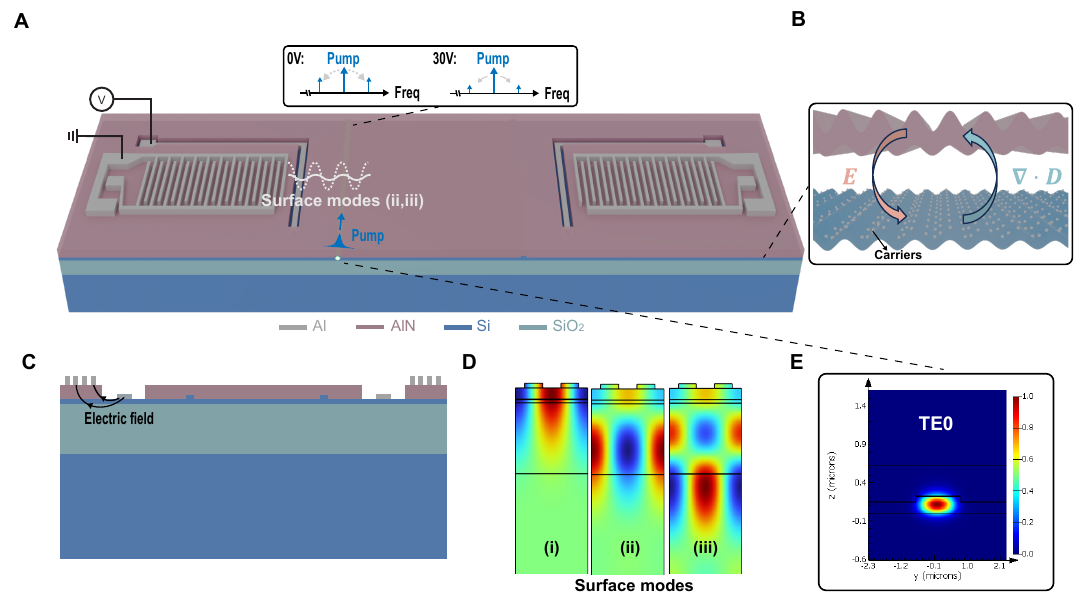} 

	\caption{\fontfamily{phv}\selectfont\footnotesize\textbf{Principles of AE modulation with acousto-optic (AO) readout. }
		(\textbf{A}) Schematic of the device structure, including an interdigital transducer (IDT) and a nearby electrode pad used to tune Sezawa and higher-order modes via the carrier-concentration-induced AE effect. An optical waveguide is patterned along the delay line, enabling optical sideband readout of the acoustically induced changes. When the DC voltage is applied, the Stokes components from non-fundamental acoustic modes are suppressed. (\textbf{B}) AE coupling between the piezoelectric layer and the semiconducting layer. (\textbf{C}) Cross-section along the acoustic delay line. The $230~\mathrm{nm}$ silicon layer sits on a $3~\mathrm{\mu m}$ oxide layer, and partial etching of the silicon forms ridge waveguides with $80~\mathrm{nm}$ ridge height. A $480~\mathrm{nm}$-thick AlN piezoelectric film is deposited on top and etched to create lower regions for electrode deposition, followed by $200~\mathrm{nm}$-thick aluminum IDTs and electrodes. The electric field between the ground pad of the IDT and the electrode pad tunes the carrier concentration beneath the IDT.  (\textbf{D}) Vertical displacement profiles of the Rayleigh mode (i), Sezawa mode (ii) and higher-order mode (iii). The profile is simulated using COMSOL Multiphysics. (\textbf{E}) Fundamental transverse electric (TE) mode supported in an AlN-coated waveguide. The profile is simulated using LUMERICAL software.
		}
	\label{fig:1} 
\end{figure*}

\noindent{\color{saRed}\fontfamily{phv}\fontseries{b}\selectfont\normalsize INTRODUCTION}\\
\noindent
The acousto-electric effect (AE) describes the coupling between free carriers and an acoustic wave in a conducting piezoelectric material, providing an electrical path to tune carrier transport dynamics and thus control acoustic modes\cite{hutson1962elastic,ridley1988space,adler2005simple}. Since a compressional acoustic wave produces spatial modulation of the carrier density and is also accompanied by a co-propagating electric field, its propagation can be tuned in two simple ways, either by changing the background carrier concentration or by applying a drift current. With the development of surface and bulk acoustic wave technologies \cite{morgan2010surface}, AE interactions have enabled a range of demonstrations, including nonreciprocal amplification in acoustic delay lines driven by drift current\cite{ghosh2019acoustoelectric,hackett2023non,malocha2018thin,hackett2021towards,du2025experimental,bhaskar2018silicon,ghosh2017nonreciprocal,mansoorzare2020acoustoelectric} and tunable dissipation in resonators via carrier density control\cite{li2024frequency,storey2021lithium} across piezoelectric-semiconductor platforms. These advances establish AE control as a versatile route to electrically reconfigurable acoustic devices and expanded functionality for microwave signal processing.\par
Beyond microwave acoustics, phonons can enable signal transduction, delay, and nonreciprocal response in photonic integrated circuits.  Through phase-matched acousto-optic (AO) coupling and Brillouin processes, phonons can be used to encode microwave signals onto light as frequency-shifted sidebands, or to mediate scattering between distinct spatial modes \cite{kittlaus2017chip,ye2025integrated,gyger2020observation}. Leveraging unique properties of acoustic waves, it is possible to bring true-time delay via the slow acoustic velocity \cite{munk2019surface}, and direction-dependent control of optical propagation to photonic circuits \cite{kittlaus2018non}. These functionalities are attracting growing interest for on-chip filters\cite{katzman2021surface}, lasers\cite{otterstrom2018silicon,chauhan2021visible}, nonreciprocal devices\cite{cheng2025terahertz}, and signal transduction\cite{zhou2024electrically,kittlaus2018rf}. Building on these concepts, the ability to combine the AE effect with photonic integrated circuits brings new strategies to electrically amplify and tune acousto-optic interactions, substantially expanding these possibilities. For example, this AE-AO combination can, in principle, enable optical amplification, reconfigurable sideband spectra, and programmable nonreciprocity. \par
Recently, theoretical frameworks have proposed using AE modulation of phonons to dynamically control Brillouin processes \cite{otterstrom2023modulation}.  However, even experimental demonstrations of AE-modulated AO scattering in piezoelectric-semiconductor platforms that do not require guided phonon modes remain challenging, as many established AE systems do not readily support low-loss optical waveguides, limiting their use in integrated microwave photonics. Hybrid platforms offer compelling solutions to this problem by combining piezoelectricity, semiconductor carrier control, and strong optical confinement within the same device region. At the same time, complex cross-sectional geometries in these multilayer structures modify piezoelectric boundary conditions, which can introduce new acousto-electric responses that broaden the range of achievable behaviors.\par
To explore how AE control can extend photonic integrated systems, we examine AE modification of elastic waves in a silicon-photonic waveguide platform.
In this work, we introduce piezoelectricity by integrating aluminum nitride on silicon-on-insulator (AlN-on-SOI) and demonstrate mode-selective AE modulation of guided phonons by doping the top silicon layer and tuning the carrier concentration beneath the interdigital transducers (IDTs). This electrical control enables up to $10~\mathrm{dB}$ (up to $20~\mathrm{dB}$ forward transmission) suppression of higher-order acoustic modes while leaving the fundamental Rayleigh mode largely unchanged; this same mode selectivity is also observed in current-induced nonreciprocity (i.e. microwave S-parameter) measurements. To interpret such mode selectivity, we unify separate results from Farnell \cite{farnell1972elastic} , Ingebrigtsen \cite{ingebrigtsen1969surface} and Gu\'eret \cite{gueret1971simple} into a two-dimensional admittance-based framework, extending beyond conventional one-dimensional models \cite{hutson1962elastic,ridley1988space,adler2005simple} and making the mode-dependent factors that govern the AE response explicit. As a first step toward AE-controlled AO operation, we pattern an optical waveguide along the acoustic delay line, which serves not only as a broadband probe of the propagating phonons, but also as a non-destructive interface between the acoustic and optical domains  \cite{kharel2016noise,zhou2024electrically}. Unlike conventional electrical readout of acoustic delay lines, this optical probing mechanism is intrinsically immune to electromagnetic interference and does not depend on electrical readout at the receiving transducer. In addition, this approach does not require acoustic confinement in phononic crystal structures or other specially engineered guided-phonon geometries, enabling a simpler and more fabrication-tolerant route to connect delay-line acoustics with integrated photonics. Finally, we demonstrate proof-of-concept multi-domain transduction and Rayleigh-mode filtering based on the mode-selective AE modulation, opening opportunities for compact, electrically reconfigurable microwave–photonic functionality in integrated piezoelectric–photonic systems within a complementary metal–oxide–semiconductor (CMOS)-compatible silicon-photonic architecture.
\begin{figure*}[!b] 
	\centering
	\includegraphics[width=0.85\textwidth]{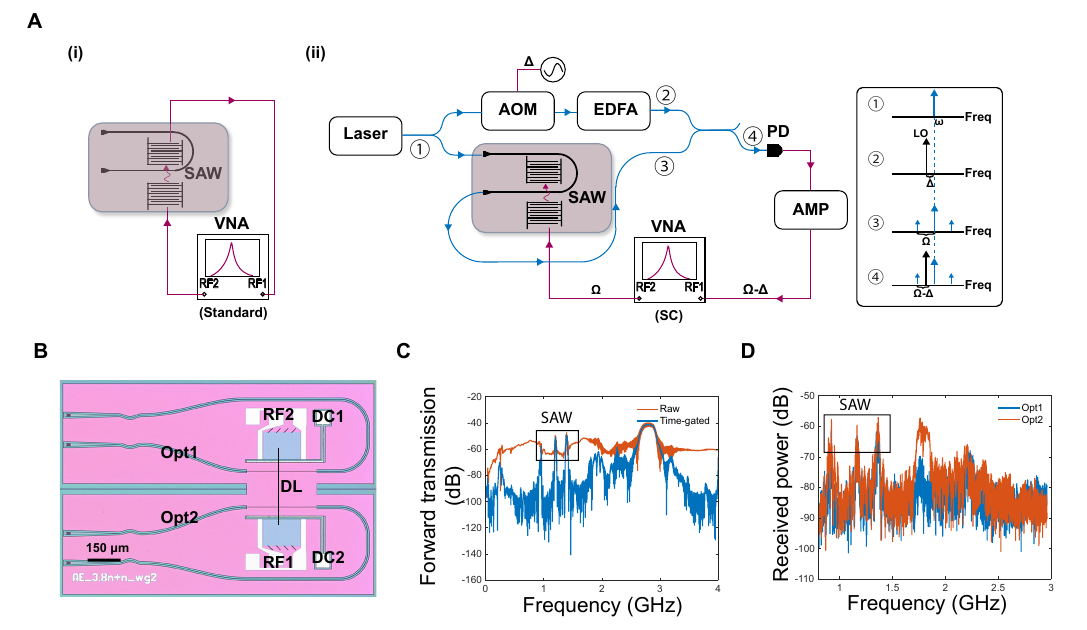} 

	\caption{\fontfamily{phv}\selectfont\footnotesize\textbf{AlN-SOI device and RF-optical readout scheme.}
		(\textbf{A}) Measurement setups. (\textbf{i}): RF S-parameter measurement using a vector network analyzer (VNA) in standard S-matrix measurement mode. (\textbf{ii}): optical heterodyne readout of optical sidebands. VNA (SC): scalar converter/mixer mode used to measure the received power at an intermediate-frequency offset ($\Delta=39~\mathrm{MHz}$) from the drive. AOM: acousto-optic modulator that shifts the optical frequency ($\Delta$)  as a local oscillator (LO). EDFA: erbium-doped fiber amplifier. PD: photodetector. AMP: RF amplifier. SAW: surface acoustic waves. Optical frequencies in each path are indicated in the right panel. (\textbf{B}) Optical micrograph of the device. RF1 and RF2 correspond to ports of the radio-frequency (RF) network; DC1 and DC2 denote the direct-current (DC) contact pads; Opt1 and Opt2 represent the optical (Opt) waveguides integrated along the acoustic delay line (DL). (\textbf{C}) Measured $\mathrm{S_{12}}$ and its time-gated spectrum. (\textbf{D}) Measured Stokes-sideband power from the heterodyne readout.
        }
	\label{fig:2} 
\end{figure*}
\vspace{2.4em}
\\
\noindent{\color{saRed}\fontfamily{phv}\fontseries{b}\selectfont\normalsize RESULTS}\\
\noindent{\fontfamily{phv}\fontseries{b}\selectfont\normalsize Operation principles}\\
\noindent
 As seen in Figure~\ref{fig:1}A, we integrate a piezoelectric AlN layer atop a silicon-on-insulator optical waveguide layer to create AE interactions in silicon photonics. The top silicon layer is doped to increase electrical conductivity, allowing the Si/AlN stack to behave as an effective conducting piezoelectric structure that supports AE interactions. In this coupling pathway, carrier dynamics induce an electric-displacement response that modifies the acoustic field and the polarization through piezoelectricity, which in turn acts back on the carriers (Figure~\ref{fig:1}B). The AE-induced spatial growth rate $\alpha$  (i.e., imaginary part of propagation constant) of the acoustic wave can be qualitatively understood using the one-dimensional theory \cite{hutson1962elastic}, yielding (Supplementary Text~\ref{sec:1})
\begin{equation}
    \alpha (\omega)=\frac{K^2 k}{2}\frac{-\frac{\omega_c}{\omega}(1-\frac{v_e}{v})}{(1-\frac{v_e}{v})^2+(\frac{\omega}{\omega_D}+\frac{\omega_c}{\omega})^2}.
    \label{eq:1}
\end{equation}
Here, $K^2$ is the piezoelectric coupling strength; $k$ is the acoustic wave number;  $\omega$ is the acoustic angular frequency; $v$ is the phase velocity of the acoustic wave and $v_e$ is the carrier drift velocity. The parameters $\omega_D$ and $\omega_c$ are effective characteristic angular frequencies that depend on the carrier concentration and diffusion. In the absence of carrier drift ($v_e=0$), Equation~(\ref{eq:1}) predicts acoustic attenuation ($\alpha<0$) whose magnitude relates to carrier concentration through $\omega_c$. When a drift current is present ($v_e\neq0$),  the spatial growth rate becomes directionally dependent. $\alpha$ can be positive for waves propagating along the current and negative for waves propagating against it, leading to acoustic nonreciprocity. To explore these effects in a silicon photonics platform, we discover a pronounced mode-selective acoustic damping enabled by carrier-density modulation, along with a modest drift-current-induced nonreciprocity (Supplementary Text~\ref{sec:15}).\par
Specifically, we pattern a DC electrode near the interdigital transducer (IDT) region to apply a lateral electric field that modulates the carrier concentration in the thin silicon layer beneath the IDTs, enabling strong damping of non-fundamental acoustic modes under bias, as shown in Figure~\ref{fig:1}C. We start with a standard photonic SOI wafer with $10~\mathrm{\Omega\cdot cm}$ p-doped silicon. To achieve Ohmic contact and reduce the voltage drop across the contact interface, the top silicon layer of this wafer is heavily n-doped only beneath DC electrodes (approximately $8\times 10^{19}~\mathrm{cm^{-3}}$, with $30~\mathrm{\Omega/\mathrm{sq}}$ sheet resistance) and more lightly n-doped ($1\times 10^{16}~\mathrm{cm^{-3}}$, with $13~\mathrm{k\Omega/\mathrm{sq}}$ sheet resistance) in the region between DC electrodes \cite{paul2020single}. The guided acoustic modes (Rayleigh, Sezawa, and higher-order surface modes; vertical displacement profiles in Figure~\ref{fig:1}D) are read out non-destructively using an optical ridge waveguide formed by partially etching the top silicon layer. This waveguide is patterned along the acoustic delay line and supports the fundamental guided optical mode shown in Figure~\ref{fig:1}E. In this process, traveling phonons scatter the pump light into Stokes and anti-Stokes sidebands with equal and opposite frequency shifts via photoelastic and moving-boundary contributions \cite{johnson2002perturbation,li2023frequency}. As a result, AE-induced tuning of non-fundamental acoustic modes is directly mapped onto changes in the measured optical sidebands.

\vspace{1.2em}
\noindent{\fontfamily{phv}\fontseries{b}\selectfont\normalsize Device structure and readout of SAW modes}\par
\noindent
Figure~\ref{fig:2}B shows an optical micrograph of the device. The IDTs set the acoustic wavelength of $3.8~\mathrm{\mu m}$, and $1550 ~\mathrm{nm}$ light is coupled on chip through grating couplers. The subsequent cosine-shaped bent section acts as a mode filter, attenuating higher-order optical modes and retaining the fundamental TE mode in the AO-active region. Two optical waveguides patterned along the acoustic delay line enable monitoring of the propagation of the acoustic modes. Each waveguide is positioned ten acoustic wavelengths away from the bias electrode to minimize interference from the scattered acoustic waves \cite{momoi1985scattering,momoi1980scattering,momoi1981scattering}. To characterize the acoustic modes, the RF S-parameters are measured using a vector network analyzer (VNA) (Figure~\ref{fig:2}A(i)). The Stokes components scattered by these acoustic modes can be measured via optical heterodyne measurement (Figure~\ref{fig:2}A(ii)), where the photodetector records the beat note between the Stokes sideband and the local oscillator (LO). Because an acousto-optic modulator (AOM) shifts the LO by a known frequency offset ($\Delta$), the beat note appears at a shifted frequency ($\Omega-\Delta$) rather than at the acoustic frequency ($\Omega$). The VNA is therefore operated in scalar mixer/converter (SC) mode to measure the converted power at this shifted frequency. \par
In Figure~\ref{fig:2}C, the forward transmission from RF2 port to RF1 port is measured. The first three peaks within the rectangular frame correspond to SAWs launched by the IDTs at $0.96 ~\mathrm{GHz}$, $1.20~\mathrm{GHz}$ and $1.39~\mathrm{GHz}$. The Rayleigh mode exhibits lower transmission than the Sezawa and higher-order modes due to stronger scattering from the AlN trench that enables contact between the bias electrode and the silicon layer. Peaks at higher frequencies are attributed to radiative bulk modes, which are excited by multilayer stack geometry and by partial reflections from the bottom boundary of the device (Supplementary Text~\ref{sec:5}). Together with the background p-doping in the silicon layer, the parasitic coupling between two RF ports induces a comparably larger electromagnetic (EM) feedthrough that reaches the receiving RF1 port without propagating in the acoustic delay line. Therefore, time gating is used to eliminate this direct port-to-port coupling, as the propagating acoustic waves have much larger group delay than the EM feedthrough. Figure~\ref{fig:2}D shows consistent behavior in the optical readout, where the Stokes-sideband spectrum exhibits three peaks associated with the SAW modes and higher-frequency features corresponding to radiative bulk waves that attenuate strongly between the two waveguides (Opt1 and Opt2).

\begin{figure*}[!b] 
	\centering
	\includegraphics[width=0.85\textwidth]{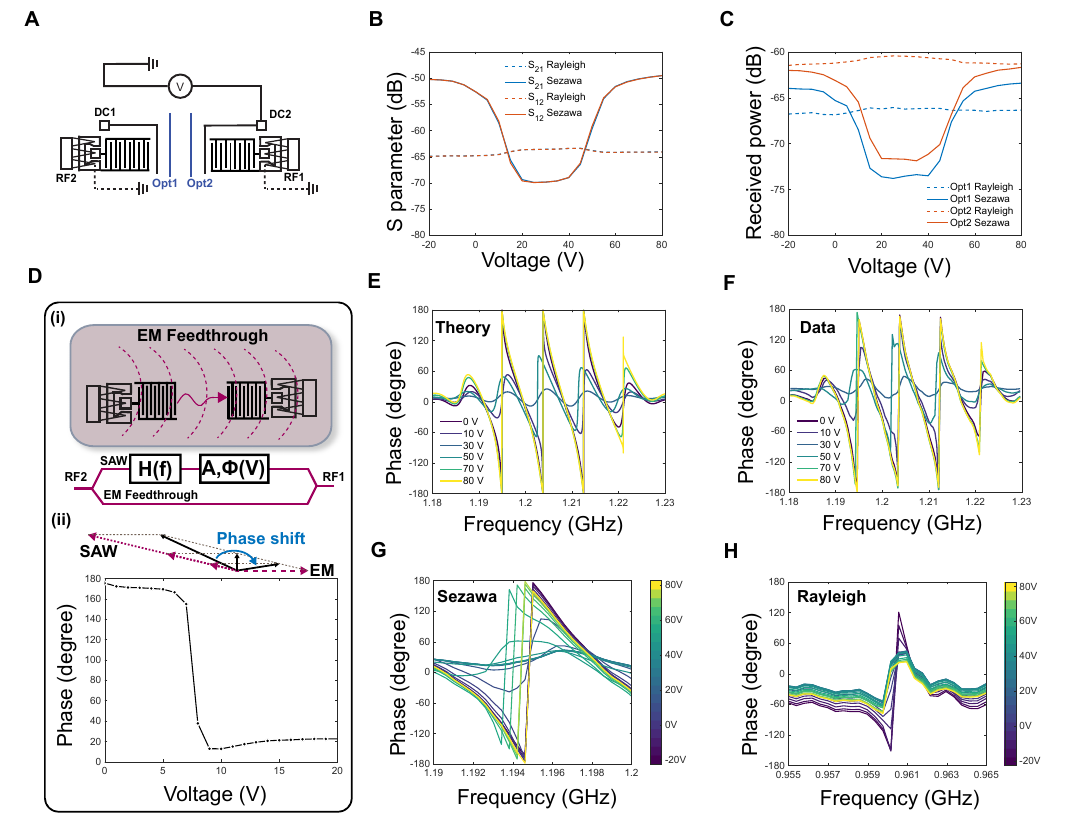} 

	\caption{\fontfamily{phv}\selectfont\footnotesize\textbf{Carrier-concentration-induced AE tuning beneath IDTs under lateral electric bias.}
		(\textbf{A}) Electrical configuration of the measurement setup. The labeling scheme is consistent with that used in Figure \ref{fig:2}B. (\textbf{B}) Time-gated S-parameter measurements under applied bias. (\textbf{C}) Optical heterodyne measurement of Stokes-sideband power under applied bias. The LO power is $-7~\mathrm{dBm}$ and the optical pump power on chip is $0~\mathrm{dBm}$. The RF power is $0~\mathrm{dBm}$. (\textbf{D}) (\textbf{i}) Interference model. From RF2 port to RF1 port, the measured signal contains contributions from an acoustic branch (IDT–SAW–IDT) and a direct electromagnetic (EM) feedthrough branch. (\textbf{ii}) Phase of $\mathrm{S_{12}}$ at $1.1946~\mathrm{GHz}$ under applied bias. The total phase shift is induced by SAW amplitude modulation and its interference with the EM feedthrough. (\textbf{E}) Phase spectrum of the Sezawa mode extracted from $\mathrm{S_{12}}$ at different voltages. (\textbf{F}) Reconstructed Sezawa-mode phase spectrum using the interference model. (\textbf{G}) Sezawa-mode phase spectrum from $\mathrm{S_{12}}$ over the full bias sweep. (\textbf{H}) Rayleigh-mode phase spectrum from $\mathrm{S_{12}}$ over the full bias sweep.
		}
	\label{fig:3} 
\end{figure*}

\vspace{1.2em}
\noindent{\fontfamily{phv}\fontseries{b}\selectfont\normalsize Selective carrier-concentration-induced AE effect}\par
\noindent
When a DC bias is applied to the electrode shown in Figure~\ref{fig:3}A, a lateral field between the electrode and the IDT ground pad tunes the surface potential of the thin silicon layer and thus modulates the carrier concentration beneath IDTs, resulting in a change of acoustic transmission through the AE effect. In the measured $\mathrm{S_{12}}$ and $\mathrm{S_{21}}$ responses (Figure~\ref{fig:3}B), the Sezawa mode is suppressed by $20~\mathrm{dB}$ with an applied voltage of $30~\mathrm{V}$ (with $16~\mathrm{dB}$ suppression for the higher-order mode), while the Rayleigh mode remains largely unchanged. This modulation is also observed through complementary AO-based phonon measurements using optical waveguides. Through optical measurements, a $10~\mathrm{dB}$ suppression of the Stokes beat note is observed at $30~\mathrm{V}$ (Figure~\ref{fig:3}C). This difference arises because the lateral field modulates the carrier concentration under both the launching and receiving IDTs. As a result, the RF delay line accumulates AE-induced damping from both the launching and receiving IDTs, doubling the suppression of the forward transmission. Since a $10~\mathrm{dB}$ suppression is observed in the IDT region with only 30 finger pairs at a $3.8~\mathrm{\mu m}$ acoustic wavelength, such a large attenuation of the Sezawa mode under the IDT region is attributed to the fact that the IDT region both forms a weakly confined acoustic resonator via partial reflection at its end and slows the acoustic group velocity through its periodic structure, increasing the effective propagation length and interaction time for carrier–acoustic energy exchange via the time-harmonic AE interaction. \par
Due to the presence of the significant EM feedthrough in our photonic platform, we next analyze the phase response of the RF delay line using the simple two-branch model shown in Figure~\ref{fig:3}D(i). The direct EM feedthrough branch does not accumulate an appreciable propagation phase between the two RF ports because the EM wavelength at gigahertz frequencies ($\sim 30~\mathrm{cm}$) is much larger than the port separation distance ($\sim 300~\mathrm{\mu m}$). As a result, the phase delay produced by feedthrough is approximately constant over the band of interest. In contrast, the SAW branch acquires both AE-induced amplitude and phase changes. The total RF response is the interference between the EM-feedthrough branch and the acoustic branch. In our AlN-SOI platform, the EM feedthrough level is approximately $-60~\mathrm{dB}$ while the Sezawa transmission is tuned from $-50~\mathrm{dB}$ to $-70~\mathrm{dB}$. When the Sezawa-mediated SAW contribution is relatively strong ($-50~\mathrm{dB}$), the measured phase is SAW-dominated and tracks the SAW phase. When it is strongly attenuated ($-70~\mathrm{dB}$), the response becomes EM-feedthrough–dominated and the measured phase instead tracks the EM phase. Within the central frequency range of the Sezawa mode, there is typically a frequency point where the EM and SAW contributions are out of phase. Sweeping the bias can shift the dominant contribution from SAW to EM at that frequency, producing an apparent phase tunability approaching $\pi$ (Figure~\ref{fig:3}D(ii)). This near-$\pi$ tunability is a distinctive signature of AE-enabled switching in a two-path interference response.\par
Under the interference model, the IDTs impose a $\mathrm{sinc^2}$ transduction response \cite{morgan2010surface}, and the AE interaction further modifies the amplitude and effective time delay. The measured transmission  between two RF ports is given by
\begin{equation}
\begin{split}
    S_{21}(\omega,V)=A_{EM}+\left|\frac{\sin(N_p(\omega-\omega_0)/\omega_0)}{N_p(\omega-\omega_0)/\omega_0}\right|^2\cdot\\
    A(V)\exp(-j\omega\tau(V)-j\omega\tau_0+j\phi_0(V)).
\end{split}
\label{eq:2}
\end{equation}
Here, $\omega$ is the angular frequency range in a narrow band around the Sezawa-mode center frequency $\omega_0$; $\tau_0$ is the RF time delay without bias; $\phi_0(V)$ is a small background phase shift. $V$ is the voltage applied; $A_{EM}$ is the amplitude of the EM feedthrough; $N_p$ is the effective number of IDT finger pairs. The AE effect tunes the amplitude of the acoustic wave $A(V)$ and the effective time delay $\tau(V)$. Using Equation~\ref{eq:2}, we use the measured Sezawa-phase response in Figure~\ref{fig:3}F to reconstruct the spectra at different voltages, as shown in Figure~\ref{fig:3}E. From the reconstruction (Supplementary Text~\ref{sec:12}), we extract an additional time delay of $0.18~\mathrm{ns}$ at $50~\mathrm{V}$ (horizontal shift along the x-axis) relative to a total RF delay of $114.16~\mathrm{ns}$ without bias, which is the result of the AE-induced phase velocity change\cite{storey2021lithium}. The phase spectra of the Sezawa and Rayleigh modes over a full voltage sweep are shown in Figure~\ref{fig:3}G and Figure~\ref{fig:3}H, respectively. Because the AE tuning predominantly affects non-fundamental modes, the Sezawa mode and higher-order modes (not shown here) exhibit a clear horizontal shift of the interference pattern of the phase spectrum.\par
We interpret such mode-selective AE responses using the two-dimensional AE model, since Equation~\ref{eq:1} does not account for mode profiles. In general, when the thin conducting sheet is buried inside the piezoelectric material, the sheet admittance $Y_s$ can be described as (Supplementary Text~\ref{sec:2})
\begin{equation}
    Y_s(k)=\frac{1}{Z_\textit{down}(k)}-\frac{1}{Z_\textit{up}(k)},
    \label{eq:3}
\end{equation}
where $k$ is the complex wave number; $Z_\textit{up}$ is the effective wave impedance seen from the top surface of the thin conducting sheet to the free piezoelectric surface; $Z_\textit{down}$ is the effective wave impedance seen from the bottom surface of the sheet into the bulk dielectric substrate. The left-hand-side sheet admittance is determined by carrier dynamics, such as those described by a drift–diffusion model, while the right-hand-side impedance terms arise from the piezoelectric boundary conditions. In the limiting case, a thin conducting sheet is on the surface of the piezoelectric material ($Z_{up}=\infty$, i.e., open-circuit), the AE response from Equation~\ref{eq:3} becomes consistent with the conventional treatment of Equation~\ref{eq:1}. For the case of a buried conducting sheet, both $Z_\textit{up}$ and $Z_\textit{down}$ are dependent on the transverse mode profiles (Supplementary Text~\ref{sec:3}). In our device, the Rayleigh mode decays exponentially from the surface into the bulk, whereas the non-fundamental modes (including radiative bulk modes) exhibit oscillatory partial waves supported by the oxide-on-silicon substrate (slow-on-fast) structure \cite{farnell1972elastic}. As a result, the AE response can differ substantially between the Rayleigh mode and the higher-order modes. 
\par
\begin{figure*}[!b] 
	\centering
	\includegraphics[width=0.85\textwidth]{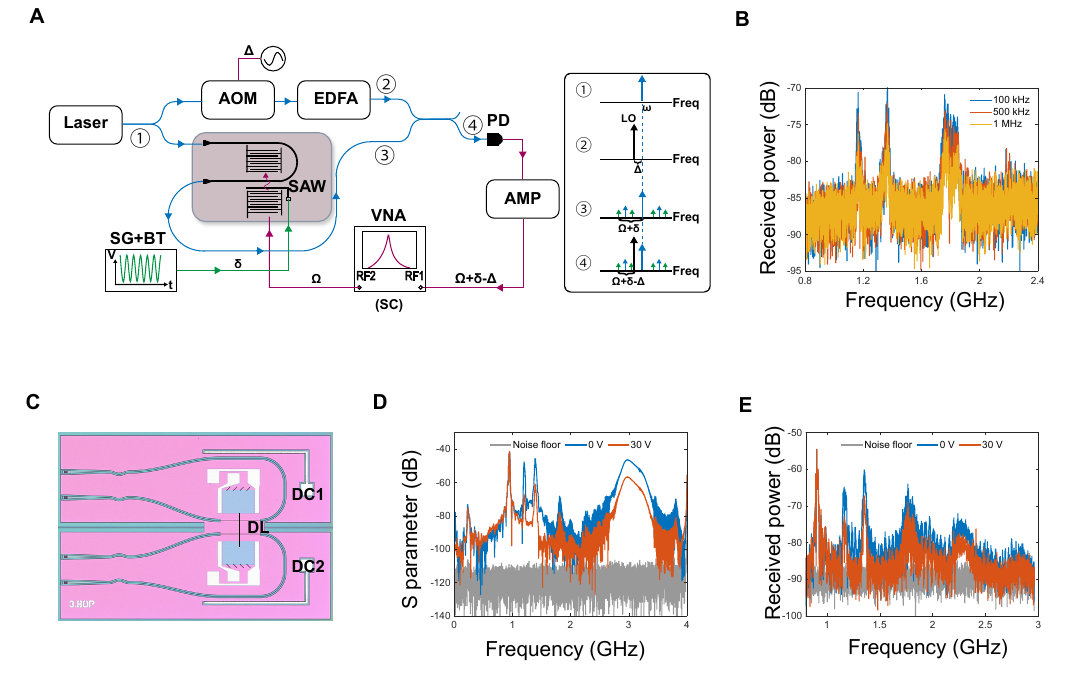} 

	\caption{\fontfamily{phv}\selectfont\footnotesize\textbf{Multi-domain signal transduction and Rayleigh-dominant filtering.}
		(\textbf{A}) Measurement setup (same as Figure~\ref{fig:2}A), with an AC modulation voltage imposed on a DC bias applied to the electrode pads to modulate the acoustic response. The beat note between the LO and the acoustically modulated optical sideband is measured using the VNA in scalar mixer/converter mode. SG: signal generator. BT: bias tee. (\textbf{B}) Measured beat-note power for different AC modulation frequencies. (\textbf{C}) Device configuration. The DC contact pads (DC1 and DC2) are located outside the delay line (DL). (\textbf{D}) Demonstration of fundamental Rayleigh-dominant filtering in the RF delay-line response. The signal is time-gated. (\textbf{E}) Demonstration of Rayleigh-dominant filtering in the optical sideband readout.}
	\label{fig:4} 
\end{figure*}
For the Rayleigh mode at $3.8~\mathrm{\mu m}$ wavelength, we observe a modest $1.2~\mathrm{dB}$ increase in transmitted power under the same bias that suppresses the higher-order modes; a similar increase is observed across different devices with IDTs of different pitches (corresponding to acoustic wavelengths from $3.8~\mathrm{\mu m}$ to $2.7~\mathrm{\mu m}$, see Supplementary Text~\ref{sec:9}). One interpretation is suggested by the two-dimensional acousto-electric model: a finite $Z_\textit{up} (k)$ term enters the sheet-admittance relation (Equation~\ref{eq:3}) and it modulates the AE interaction in a more complicated form. This added complexity is also consistent with mode-decomposition perspectives. Behunin \textit{et al.} model the AE response as the sum of contributions from multiple surface-plasmon eigenmodes coupled to a given acoustic mode, where each contribution is weighted by a coupling strength proportional to the plasmon–phonon mode overlap \cite{behunin2026noise}. An applied DC bias modifies the surface potential of the conducting sheet, which in turn changes the boundary conditions that define these plasmon eigenmodes. As a result, the coupling strengths of different plasmon eigenmodes are modified unequally, and the net AE response in the buried conducting-sheet case need not remain purely dip-like as the voltage is swept, and may instead exhibit a weak bump. Since AE gain is possible only when there is drift current, we interpret the appearance of such a bump as reduced acoustic damping rather than the true acoustic gain. 
\par
A well-known characteristic of the AE interaction is its non-reciprocal nature. According to Equation~\ref{eq:1} of the main text, the carrier drift can either supply energy to, or extract energy from, the acoustic wave depending on the relative velocity between the carriers and the acoustic wave. If the carrier velocity is larger than the acoustic velocity while they co-propagate, the acoustic wave is effectively dragged by the carriers and exhibits gain (a positive spatial growth rate). If the carrier velocity is smaller, or directed opposite to the acoustic propagation, the interaction instead introduces extra loss. Therefore, under an applied current along the delay line, the transmission for forward and backward acoustic propagation relative to the carrier velocity is different, leading to non-reciprocity that can be measured in S-parameters. As an important verification of this mode-selective AE behavior, we also measure the current-induced non-reciprocal response (Supplementary Text), which shows the consistent, selective current-induced AE response that gives rise to the non-reciprocity (the non-reciprocal transmission ratio is $4.7~\mathrm{dB/mm}$).

\vspace{1.2em}
\noindent{\fontfamily{phv}\fontseries{b}\selectfont\normalsize Multi-domain transduction and Rayleigh-dominant filtering}\par
\noindent
To utilize the mode-selective response, the electrical control of non-fundamental acoustic modes enables multi-domain signal transduction by modulating these modes and optically reading out the modulation through AO interactions. An on-chip AC electrical signal at an appropriate DC bias (which sets the working point) can be transduced into the acoustic (and thus RF) domain, and the resulting modulated acoustic wave is further converted into optical sidebands. We established such a multi-domain modulation setup in Figure~\ref{fig:4}A. A signal generator provides a $V_{pp}=2~\mathrm{V}$ AC electric signal with frequency $\delta$ and a $3~\mathrm{V}$ DC bias is supplied by a DC source (not shown). These signals are combined using a bias tee and applied to the electrode to dynamically modulate the carrier concentration beneath the IDT regions and generate acoustic sidebands. All other equipment remains the same as in Figure~\ref{fig:2}A. In the setup, the LO power is $4~\mathrm{dBm}$, the optical pump's power on chip is $8~\mathrm{dBm}$ and the input RF power is $14~\mathrm{dBm}$. The VNA then measures the received beat-note power (with frequency $\Omega+\delta-\Delta$) between the acoustic sideband and the LO. Figure \ref{fig:4}B shows the received power spectrum from the Opt2 waveguide under different AC electrical modulation frequencies ($\delta$) and indicates an approximate $500~\mathrm{kHz}$ 3-dB modulation bandwidth, mainly constrained by the equivalent RC time constant of the electrical loop. In this spectrum, three peaks corresponding to the Sezawa mode, the higher-order surface mode, and the bulk mode are observed since these modes are AE-active. Because the Rayleigh mode does not show a pronounced AE response, there is no corresponding modulation peak. Such carrier-concentration-induced AE modulation can be useful in sensing applications. Some SAW sensors encode the measurand as a quasi-static phase shift \cite{schmalz2020multi,pan2024passive,vanotti2013saw}, where direct phase readout can be vulnerable to background noise and parasitic signals such as electromagnetic feedthrough. In principle, on-chip AE modulation could tag the acoustic branch only, while leaving the electromagnetic feedthrough largely unmodulated. Combined with nondestructive optical probing of the localized acoustic response, detecting the resulting modulation sideband could provide a lock-in-style readout channel that is less sensitive to low-frequency background noise and electromagnetic feedthrough in both RF domain and optical domain, enabling potentially a more robust readout.\par
Since the AE modulation predominantly occurs in non-fundamental modes, the strong suppression of these modes allows a Rayleigh-dominant response, which is advantageous for applications requiring well-defined Rayleigh-mode propagation on the chip. To avoid the scattering of the Rayleigh mode along the delay line, we test a device in which the electrodes are patterned only outside the IDT–delay-line–IDT region, as shown in Figure~\ref{fig:4}C. The remaining electric configuration is identical to that of Figure~\ref{fig:3}A. With an applied bias, the device exhibits a cleaner Rayleigh response ($> 20~\mathrm{dB}$ rejection of other passbands compared with the Rayleigh passband in Figure~\ref{fig:4}D). This selectivity is useful for broadband excitation schemes such as wireless electromagnetic coupling\cite{pan2024passive} and chirped IDTs\cite{fall2017generation}, the unwanted passbands, such as the Sezawa passband, can create hard-to-predict interference and noise along the delay line. For nominally narrowband excitation cases, the modest Rayleigh-Sezawa frequency difference can still permit Sezawa sidelobes to leak into the Rayleigh band. Therefore, suppressing these contributions to obtain a Rayleigh-only response produces a cleaner transfer function and time-domain impulse response. Similar behavior is observed in Figure~\ref{fig:4}E with optical sideband readout. Although Sezawa-mode suppression is weaker optically than in RF, additional rejection can be engineered geometrically by preferentially enhancing AO coupling to the fundamental Rayleigh mode while reducing coupling to higher-order modes. For example, because higher-order modes are largely confined to the oxide buffer layer, suspended structures may further suppress them. This mode selectivity is thus attractive not only for RF signal processing, but also for photonic implementations such as multiplexing, where multiple IDTs at different RF carrier frequencies share a single optical waveguide readout. Ensuring a Rayleigh-only response reduces cross-talk and avoids ambiguous spectral assignment. More generally, suppressing higher-order modes can function as an electrically controlled on/off gate for optical sideband generation, offering a building block for reconfigurable microwave–photonic signal processing.

\vspace{2.4em}
\noindent{\color{saRed}\fontfamily{phv}\fontseries{b}\selectfont\normalsize DISCUSSION}\\
\noindent
Building on the results in the AlN–SOI platform, which already suggest useful mode selectivity, several directions could further increase carrier-drift-induced AE gain to boost AO scattering in photonic platforms. First, employing piezoelectric films with larger electromechanical coupling $K^2$ directly enhances the AE interactions. For instance, Aluminum Scandium nitride (AlScN) is a CMOS-compatible material with reported $K^2>2\%$ \cite{wang2014high,kobayashi2021high}. Second, the doping strategy should be carefully managed to maximize AE performance. Third, acoustic mode engineering remains critical, since the spatial mode profile directly sets the coupling strength and selectivity of the AE interaction. \par
In summary, we observed the mode-selective AE effect in a scalable AlN-SOI platform, achieving up to $20~\mathrm{dB}$ forward transmission suppression of higher-order modes across the acoustic delay line. This strong AE damping under the IDT regions enables a unique near-$\pi$ phase tuning under bias. Meanwhile, the fundamental Rayleigh mode remains largely unchanged. We explain the mode selectivity using an admittance-matching picture, which introduces geometry-dependent factors not captured by conventional AE treatments for the buried conducting sheet. Building on this large suppression and selectivity, we demonstrate multi-domain signal transduction and a Rayleigh-dominant filtering response, pointing to new strategies for reconfigurable mode engineering and microwave-photonic functionalities.

\vspace{2.4em}
\noindent{\color{saRed}\fontfamily{phv}\fontseries{b}\selectfont\normalsize MATERIALS AND METHODS}\par

\noindent{\fontfamily{phv}\fontseries{b}\selectfont\normalsize Fabrication of the device}\par

\noindent
These silicon photonic devices with mode-selective acousto-electric modulation are fabricated on single-crystal silicon-on-insulator (SOI) wafers using Sandia’s Microsystems Engineering, Science and Applications (MESA) CMOS production and research facilities. The features in the silicon device layer, oxide, aluminum nitride (AlN), and aluminum (Al) metal are patterned and defined with deep-UV photolithography and plasma etching. After processing the optical waveguide structures and isolation trenches in the silicon device layer, the acoustic delay line regions in the silicon are created with a phosphorous implantation step to produce regions of n doped silicon with a sheet resistance $\sim13 ~\mathrm{k\Omega/sq}$. Subsequently, the Ohmic contact regions in the silicon are created with a second phosphorous implantation step to produce regions of n+ doped silicon with a sheet resistance $\sim30~\mathrm{\Omega/sq}$. We then deposit $800~\mathrm{nm}$ of plasma-enhanced chemical-vapor-deposited (PECVD) oxide and employ chemical mechanical polishing (CMP) to thin the oxide to the desired thickness above the silicon device layer ($\sim300~\mathrm{nm}$). This planarized oxide layer serves as an etch stop for the subsequent AlN and Al layers, which also protects the optical and acoustic structures in the silicon device layer during these etch steps. After the oxide layer is etched in the Ohmic contact and acousto-electric regions to expose the silicon device layer, a $480~\mathrm{nm}$ AlN piezoelectric thin film is deposited using RF sputtering such that the c-axis of the crystal grains is oriented normal to the wafer. After lithographic patterning and etching of the AlN thin film, a $200~\mathrm{nm}$ Al metal film is deposited, patterned, and etched to define the device electrodes. 
\vspace{1.2em}\par
\noindent{\fontfamily{phv}\fontseries{b}\selectfont\normalsize RF measurement}\par
\noindent
Prior to measurement, the VNA (KEYSIGHT P5004A) is calibrated to the GSG probe tips (PrOTEQ SOLUTIONS) using on-wafer short/open standards. A source meter (KEITHLEY 2400) provides the DC bias, with one terminal tied to the VNA ground. The S parameters from $20~\mathrm{MHz}$ to $4~\mathrm{GHz}$ (the frequency window is large enough to perform the inverse Fourier transform) are acquired twice at each bias point and averaged. During the sweep, the current is sampled every $0.5~\mathrm{s}$ and the time-averaged current was recorded. The IF bandwidth is set to be $3~\mathrm{kHz}$ and the number of measured points across the frequency range is $10000$. All instrument control and data recording are automated using Python scripts.
\vspace{1.2em}\par
\noindent{\fontfamily{phv}\fontseries{b}\selectfont\normalsize Optical and modulation measurement}\par

\noindent
The pump's optical wave is provided by an ORION $1550~\mathrm{nm}$ laser. The LO branch is composed of a $39~\mathrm{MHz}$ AOM (BRIMROSE) and an EDFA (AGILTRON). A polarizer is placed in each branch to align polarization and maximize the constructive interference. The PD (ORTEL 2860D-C04) has a responsivity of $0.85~\mathrm{A/W}$ with $500~\mathrm{\Omega}$ transimpedance and a $10~\mathrm{GHz}$ bandwidth, followed by a low noise RF amplifier (ZX60-83LN-S+ from MINICIRCUIT) with $20~\mathrm{dB}$ gain. At each bias, the beat note between the Stokes spectrum and the LO for the optical measurement is measured five times. The VNA is operated at the scalar/mixer converter mode ($39~\mathrm{MHz}$ mixer frequency) with the sweeping frequency from $800~\mathrm{MHz}$ to $3~\mathrm{GHz}$. In the modulation measurement, the AC electrical signal is generated by a function generator (RIGOL) and the VNA mixer frequency is changed based on the frequency of the AC electrical signal. The IF bandwidth is set to be $3~\mathrm{kHz}$ and the number of measured points across the frequency range is $4000$. All instrument control and data recording are automated using Python scripts.
\vspace{1.2em}\par
\noindent{\fontfamily{phv}\fontseries{b}\selectfont\normalsize Key results from the two-dimensional acousto-electric model}\par

\noindent
Results from Farnell \cite{farnell1972elastic} , Ingebrigtsen \cite{ingebrigtsen1969surface} and Gu\'eret \cite{gueret1971simple} can be unified into a two-dimensional admittance-based framework to interpret the mode selectivity. Under an applied external field, the carriers (here we consider electrons) in the semiconductor layer obey the drift-diffusion equation, together with current continuity and Gauss's law, leading to the sheet impedance (Supplementary Text)
\begin{equation}
    Z_s=\frac{-i}{v\varepsilon}(kh(-1+\frac{-i\omega_c/\omega }{1 +\mu fE_{DC}/v+i\frac{\omega}{\omega_D}}))^{-1}, 
    \label{eq:4}
\end{equation}
where 
\begin{equation}
    \omega_c=\frac{n|q|\mu}{\varepsilon},\omega_D=\frac{|q|\beta}{f\mu}v^2, v=\frac{\omega}{k}.
    \label{eq:5}
\end{equation}
In the above equations, $k$ and $\omega$ are the wave number and angular frequency of an acoustic wave; $q$ is the charge of the electron; $v$ is the phase velocity of acoustic waves; $\mu$ is the electron mobility; $f$ is the fraction of charge in the conduction band among all space charges; $h$, $n$ and $\varepsilon$ are the thickness, carrier concentration and permittivity of the semiconductor layer; $\beta$ is the inverse thermal energy; $\omega_c$ is the drift-related characteristic angular frequency and $\omega_D$ is the diffusion-related characteristic angular frequency; $Z_s$ is the semiconducting sheet impedance (For forward propagation, we use the convention $\exp(ikx-i\omega t)$). When the semiconductor layer is on top of the geometry, Equation \ref{eq:4} becomes
\begin{equation}
    Z_s(k)=Z_\textit{down}(k),
    \label{eq:6}
\end{equation}
where $Z_\textit{down}$ is the effective wave impedance seen from the free surface of the geometry into the bulk dielectric substrate, assuming the sheet is thin enough to neglect its mass loading. The free mechanical boundary conditions and the electric displacement continuity determine the effective wave impedance
\begin{equation}
    Z_{\textit{down}}=\frac{-i}{v\varepsilon}\frac{k_0-k}{k_\infty-k},
    \label{eq:7}
\end{equation}
where $k_0$ is the short-circuit wave number and $k_\infty$ is the open-circuit wave number. From Equation \ref{eq:4}, Equation \ref{eq:6} and Equation \ref{eq:7}, the acoustic spatial growth rate is
\begin{equation}
    \alpha=\frac{-\frac{\omega_c}{\omega}(1-\frac{fv_e}{v})k'(0) h}{(1-\frac{fv_e}{v})^2+\frac{\omega}{\omega_D}(\frac{\omega}{\omega_D}+\frac{\omega_c}{\omega})+\gamma(\frac{\omega}{\omega_D}+\frac{\omega_c}{\omega})^2},
    \label{eq:8}
\end{equation}
where $v_e$ is the drift velocity of electrons; $k'(0)$ is the slope of the wave-number-vs-sheet-thickness curve at zero sheet thickness; $\gamma$ is a parameter that is equal to $2kh-k_0h$.  Noting that $k'(0)h<k_0-k_\infty=K^2k_\infty/2$, the above equation is consistent with the one-dimensional theory of Equation~\ref{eq:1}, except that the carrier coupled terms in denominator and the scaling factor in numerator are modified. However, when the semiconductor sheet is buried inside the geometry at depth $x_3$, the acoustic behaviors become complicated and the mode selectivity AE interaction becomes significant, deviating the AE interaction predicted from Equation \ref{eq:1} and Equation \ref{eq:8}. On the one hand, the effective wave impedance seen at this depth into the bulk dielectric substrate is
\begin{equation}
    Z_{\textit{down}}=\frac{\omega}{iv^2}\frac{\phi_s(x_3)}{D_{3s}(x_3)},
    \label{eq:9}
\end{equation}
where $\phi_s(x_3)$ is the potential at depth $x_3$ and $D_{3s}$ is the vertical component of the electric displacement at this depth. For hexagonal piezoelectric materials (similar results for other classes), the potential and the vertical electric displacement can be described using partial-wave expansion
\begin{equation}
    \phi_s(x_3)=\sum_{n=1}^3C_n\alpha_4^{(n)}\exp(ikb^{(n)}x_3)\exp(ik(x-vt)),
    \label{eq:10}
\end{equation}
and
\begin{equation}
    D_{3s}(x_3)=\sum_{n=1}^3C_nD_3^{(n)}\exp(ikb^{(n)}x_3)\exp(ik(x-vt)),
    \label{eq:11}
\end{equation}
where $\alpha_4^{(n)},C_n,D_3^{(n)}$ are related coefficients and the complex number $b^{(n)}$ determines transverse profile of acoustic mode when the semiconducting layer is no longer at the surface, which is very different between the fundamental Rayleigh mode that decays into the bulk and the higher-order modes that oscillate along the transverse direction. On the other hand, finite impedance $Z_{\textit{up}}$ further modifies the AE interaction. Both factors contribute to the AE mode selectivity. 
\vspace{2.4em}\par
\noindent{\color{saRed}\fontfamily{phv}\fontseries{b}\selectfont\normalsize Supplementary Materials}\\
\noindent{\fontfamily{phv}\fontseries{b}\selectfont\footnotesize This PDF file includes:}\par
\noindent
\fontfamily{phv}\selectfont\footnotesize Supplementary Text\\
\fontfamily{phv}\selectfont\footnotesize Figures S1 to S10\\
\fontfamily{phv}\selectfont\footnotesize References
\par
\vspace{2.4em}
\noindent{\color{saRed}\fontfamily{phv}\fontseries{b}\selectfont\normalsize REFERENCES}\par
\renewcommand\refname{\vspace{-2\baselineskip}}
\footnotesize
\bibliography{main_reference} 

\vspace{2.4em}
{\fontfamily{phv}\selectfont\footnotesize
\noindent\textbf{Acknowledgments:}We thank B.Li, Y.Luo, H.H.Diamandi, and T.Cigeroglu for technical assistance with the experimental setup and helpful discussion of acousto-electric and acousto-optic results. \textbf{Funding:} This material is based upon work supported {}by the Laboratory Directed Research and Development program at Sandia National Laboratories. Sandia National Laboratories is a multiprogram laboratory managed and operated by National Technology and Engineering Solutions of Sandia, LLC, a wholly owned subsidiary of Honeywell International, Inc., for the U.S. Department of Energy’s National Nuclear Security Administration under Contract No. DE-NA-0003525. This paper describes objective technical results and analyses. The views, opinions, and/or findings expressed are those of the authors and should not be interpreted as representing the official views or policies of the U.S. Department of Energy, U.S. Department of Defense, or the U.S. Government. Part of the research was carried out at the Jet Propulsion Laboratory, California Institute of Technology, under a contract with the National Aeronautics and Space Administration. This research was developed with funding Sandia's LDRD and the National Science Foundation (NSF) under Grant No. 2137740. \textbf{Author contributions:} Y.Z., N.T.O., M.J.S and P.T.R conceived and planned the experiments. R.Y. carried out the acousto-electric and acousto-optic experiments, analytical derivations and data analysis with assistance from N.T.O., M.J.S, R.O.B, H.C., B.S., N.T.O., M.J.S., D.C.T. and M.E. contributed to the interpretation of the results. M.J.S., A.L.S., A.L.L. and N.T.O. fabricated the devices. R.Y. and P.T.R. led the manuscript writing. All authors provided critical feedback to shape the research, analysis and manuscript. \textbf{Competing interests:} P.T.R. is a founder and shareholder of Resonance Micro Technologies Inc. \textbf{Data, code, and materials availability:} All data and code needed to evaluate and reproduce the results in the paper are present in the paper and/or the Supplementary Materials. The original data of the paper will deposit in Zenodo digital repository after publication (DOI: 10.5281/zenodo.18856636). This study did not generate new materials.
\label{LastMainPage}

\newpage


\renewcommand{\thefigure}{S\arabic{figure}}
\renewcommand{\thetable}{S\arabic{table}}
\renewcommand{\theequation}{S\arabic{equation}}
\renewcommand{\thepage}{\arabic{page}}
\setcounter{figure}{0}
\setcounter{table}{0}
\setcounter{equation}{0}


\clearpage
\setcounter{page}{1}
\fancyfoot{}
\fancyfoot[C]{\raisebox{-2.5em}{\hspace*{-6em}\textbf{\thepage\ of \pageref{LastPage}}}}
\newgeometry{margin=1in}
\fontfamily{ptm}\fontsize{12}{14}\selectfont
\onecolumn
\linespread{1.5}\selectfont
\makeatletter
\renewcommand{\fnum@figure}{\textbf{Fig. \thefigure}}
\makeatother
\begin{center}
\section*{Supplementary Materials for\\ \scititle}

Ruoyu~Yuan,
Yishu~Zhou,
Matthew~J.~Storey,
Ryan~O.~Behunin, \\
Haotian~Cheng,
Betul~Sen,
Andrew~L.~Starbuck,
Douglas~C.~Trotter,\\
Andrew~L.~Leenheer,
Matt~Eichenfield,
Nils~T.~Otterstrom,
Peter~T.~Rakich$^{\ast}$\\ 
\small$^\ast$Corresponding author. Email: peter.rakich@yale.edu\\
\end{center}

\subsubsection*{\large This PDF file includes:}
Materials and Methods\\
Supplementary Text\\
Figures S1 to S11\\

\newpage


\makeatletter
\renewcommand\subsection{\@startsection{subsection}{2}{\z@}%
  {-3.25ex\@plus -1ex \@minus -.2ex}%
  {1.5ex \@plus .2ex}%
  {\normalfont\Large\bfseries}}

\renewcommand\subsubsection{\@startsection{subsubsection}{3}{\z@}%
  {-3.25ex\@plus -1ex \@minus -.2ex}%
  {1.5ex \@plus .2ex}%
  {\normalfont\Large\bfseries}}
\makeatother

\subsection*{Supplementary Text}
\setcounter{secnumdepth}{3}
\renewcommand{\thesubsubsection}{S\arabic{subsubsection}}
\subsubsection{One-dimensional acousto-electric model}
\label{sec:1}
\noindent
The interaction between acoustic waves and carriers in the piezoelectric material is commonly described as a one-dimensional model \cite{hutson1962elastic}. The piezoelectric material is doped to a free-carrier concentration $n$. When a longitudinal acoustic wave is present, the resulting volumetric deformation of the piezoelectric material produces a small traveling carrier-density perturbation $n_s$, and the piezoelectricity generates the small traveling electric field $E$. When an external DC electric field $E_\text{DC}$ is applied, carriers (assume electrons) follow the drift-diffusion equation
\begin{equation}
    J=-q(n+fn_s)\mu (E+E_{DC})+\frac{\mu}{\beta}f\frac{\partial n_s}{\partial x},
    \label{eq:S1}
\end{equation}
where $q$ is defined as a negative value for electrons; $\mu$ is the mobility of the electron; $\beta$ is $1/k_BT$; $k_B$ is the Boltzmann constant; $T$ is the temperature; $f$ is a fraction factor of the charge in the conduction band among all space charges. In one dimension, current continuity is given by
\begin{equation}
    \frac{\partial J}{\partial x}=-q\frac{\partial n_s}{\partial t}
    \label{eq:S2},
\end{equation}
and Gauss's law is given by
\begin{equation}
    \frac{\partial D}{\partial x}=qn_s,
    \label{eq:S3}
\end{equation}
where $D$ is the electric displacement. These three equations couple the free-carrier domain and the the electrostatic domain. To solve them, expand the drift-diffusion equation and take the derivative
\begin{equation}
    \frac{\partial J}{\partial x}=-q\mu n\frac{\partial E}{\partial x}-q\mu f(\frac{\partial n_s}{\partial x}E+n_s\frac{\partial E}{\partial x})-q\mu fE_{DC}\frac{\partial n_s}{\partial x}+\frac{\mu}{\beta}f\frac{\partial^2 n_s}{\partial x^2}.
    \label{eq:S4}
\end{equation}
Substitute the current density term in the current continuity equation and take the time derivative
\begin{equation}
    \frac{\partial^2D}{\partial x\partial t}=q\mu n\frac{\partial E}{\partial x}+q\mu f(\frac{1}{q}\frac{\partial^2 D}{\partial x^2}E+\frac{1}{q}\frac{\partial D}{\partial x}\frac{\partial E}{\partial x})+\mu f E_{DC}\frac{\partial^2 D}{\partial x^2}-\frac{\mu f}{\beta q}\frac{\partial^3 D}{\partial x^3}.
    \label{eq:S5}
\end{equation}
Since the electric field and the electric displacement are attached to the traveling acoustic wave in the piezoelectric material, the plane traveling wave assumes
\begin{equation}
    D=D_0\exp(i(kx-\omega t)), E =E_0\exp(i(kx-\omega t))
    \label{eq:S6},
\end{equation}
where $k$ and $\omega$ represent the acoustic wave number and the angular frequency. At steady state, the result is
\begin{equation}
    k\omega D_0=ikq\mu nE_0+\mu f(-k^2D_0E_0-k^2D_0E_0)+\mu fE_{DC}(-k^2)D_0-\frac{\mu f}{\beta q}(-ik^3)D.
    \label{eq:S7}
\end{equation}
Rearrange the above equation
\begin{equation}
    (-\omega-2\mu fkE_0+\frac{\mu fk^2i}{\beta q}-\mu fE_{DC}k)D_0=-q\mu n E_0i.
    \label{eq:S8}
\end{equation}
Note that the term $2\mu fkE_0$ can be neglected since it is originally from the higher order term $n_sE$, we get
\begin{equation}
    D_0=\frac{q\mu n E_0i/\omega}{1-\frac{\mu fk^2i}{\beta q\omega}+\mu fE_{DC}k/\omega}.
    \label{eq:S9}
\end{equation}
The acoustic mode is assumed to be non-dispersive with phase velocity $v$, yielding
\begin{equation}
    D_0=\frac{-\omega_c\varepsilon E_0i/\omega}{1+\mu fE_{DC}/v+i\frac{\omega}{\omega_D}},
    \label{eq:S10}
\end{equation}
where we define parameters for simplicity
\begin{equation}
    \omega_c=\frac{n|q|\mu}{\varepsilon}, \omega_D=\frac{|q|\beta}{f\mu}v^2.
    \label{eq:S11}
\end{equation}
From the one dimensional piezoelectric coupling 
\begin{equation}
    D=eS+\varepsilon E,
    \label{eq:S12}
\end{equation}
where $e$ is the piezoelectric constant and $\varepsilon$ is the static permittivity of the material. The strain field $S$ is assumed to be plane traveling wave
\begin{equation}
    S=S_0\exp(i(kx-\omega t)).
    \label{eq:S13}
\end{equation}
The relation between the strain and the electric field becomes
\begin{equation}
    \frac{-eS_0}{\varepsilon}=(1-\frac{-\omega_c i/\omega}{1+\mu fE_{DC}/v+i\frac{\omega}{\omega_D}})E_0.
    \label{eq:S14}
\end{equation}
Since 
\begin{equation}
    T=cS-eE,
    \label{eq:S15}
\end{equation}
where $T$ is the stress and $c$ is the stiffness constant. By combining the above equations, the coupling between the acoustic wave and the free carriers determines the effective stiffness constant
\begin{equation}
    T_0=c^{eff}S_0,
    \label{eq:S16}
\end{equation}
where
\begin{equation}
    c^{eff}=c+\frac{e^2/\varepsilon(1+\frac{\mu fE_{DC}}{v}+i\frac{\omega}{\omega_D})}{1+\frac{\mu fE_{DC}}{v}+i(\frac{\omega}{\omega_D}+\frac{\omega_c}{\omega})}.
    \label{eq:S17}
\end{equation}
The acoustic wave needs to satisfy the Christoffel's equation in one dimension
\begin{equation}
    \rho \omega^2=c^{eff}k^2.
    \label{eq:S18}
\end{equation}
Due to the modification of the stiffness constant, we replace the $k$ vector with
\begin{equation}
    k\to k-i\alpha,
    \label{eq:S19}
\end{equation}
where $\alpha$ is defined as the gain coefficient. Under small damping/gain
\begin{equation}
    (k-i\alpha)^2\approx k^2-2i\alpha k.
    \label{eq:S20}
\end{equation}
The effective stiffness can be written as the complex number
\begin{equation}
    c^{eff}=R+Ii.
    \label{eq:S21}
\end{equation}
In the Christoffel's equation, since the left-side term is real, the imaginary part on the right side should vanish 
\begin{equation}
    -2R\alpha i+Iki=0,
    \label{eq:S22}
\end{equation}
which is
\begin{equation}
    \alpha=\frac{Ik}{2R}.
    \label{eq:S23}
\end{equation}
Form the effective stiffness constant
\begin{equation}
    \frac{I}{R}=\frac{-e^2/\varepsilon\frac{\omega_c}{\omega}(1+\frac{\mu fE_{DC}}{v})}{c((1+\frac{\mu fE_{DC}}{v})^2+(\frac{\omega}{\omega_D}+\frac{\omega_c}{\omega})^2)+e^2/\varepsilon((1+\frac{\mu fE_{DC}}{v})^2+\frac{\omega}{\omega_D}(\frac{\omega}{\omega_D}+\frac{\omega_c}{\omega}))}.
    \label{eq:S24}
\end{equation}
Note that
\begin{equation}
    \frac{e^2}{c\varepsilon}=\frac{K^2}{1-K^2}\approx K^2\ll 1,
    \label{eq:S25}
\end{equation}
where $K^2$ is the well-known electromagnetic coupling coefficient. The gain coefficient is
\begin{equation}
    \alpha=\frac{K^2 k}{2}\frac{-\frac{\omega_c}{\omega}(1+\frac{\mu fE_{DC}}{v})}{(1+\frac{\mu fE_{DC}}{v})^2+(\frac{\omega}{\omega_D}+\frac{\omega_c}{\omega})^2}.
    \label{eq:S26}
\end{equation}
Since
\begin{equation}
    v_e=-\mu E_\text{DC},
    \label{eq:S27}
\end{equation}
the gain coefficient becomes
\begin{equation}
    \alpha=\frac{K^2 k}{2}\frac{-\frac{\omega_c}{\omega}(1-\frac{fv_e}{v})}{(1-\frac{fv_e}{v})^2+(\frac{\omega}{\omega_D}+\frac{\omega_c}{\omega})^2}.
    \label{eq:S28}
\end{equation}
The same final expression holds for hole-type carriers. A special case arises when the net carrier drift velocity is zero
\begin{equation}
    \alpha=\frac{K^2 k}{2}\frac{-\frac{\omega_c}{\omega}}{1+(\frac{\omega}{\omega_D}+\frac{\omega_c}{\omega})^2}
    \label{eq:S29}.
\end{equation}
We can choose representative parameters to build intuition for the acousto-electric effect in the one-dimensional model. In Fig.~\ref{fig:S1}(A), the acoustic wave extracts energy from the drift current when its propagation velocity is lower than the carrier drift velocity (light blue region), whereas it transfers energy to the carriers when it propagates faster than the carrier drift (light red region). This asymmetry leads to nonreciprocity when a drift current is applied to the acoustic delay line. The carrier concentration also tunes the acoustic gain coefficient, as shown in Fig.~\ref{fig:S1}(B). During a linear sweep of the electrostatic potential, the carrier concentration changes exponentially, producing a dip-shaped damping feature in the acoustic response.
\begin{figure}[!b] 
	\centering
  	\includegraphics[width=0.8\textwidth]{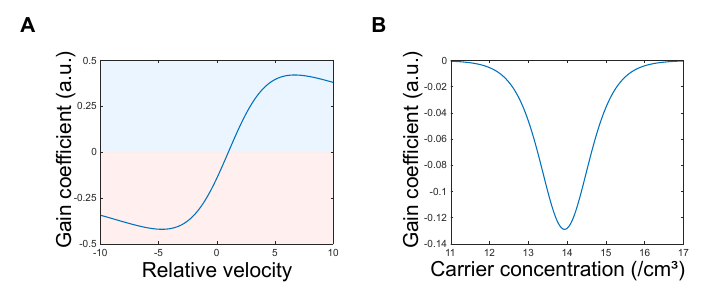} 
	\caption{\textbf{An example of the acousto-electric effect.}
		The donor concentration is $10^{14} ~\mathrm{cm^{-3}}$ and the mobility is $0.1 ~\mathrm{m^2/(s\cdot V)}$. The static dielectric permittivity is $12$, and the velocity of the acoustic mode is $3500~\mathrm{m/s}$. The acoustic wave frequency is $1~\mathrm{GHz}$. (A) The gain coefficient is controlled by the drift-carrier velocity relative to the acoustic phase velocity. (B) The gain coefficient is controlled by the carrier concentration (log scale).
		}
	\label{fig:S1} 
\end{figure}
\setcounter{secnumdepth}{3}
\renewcommand{\thesubsubsection}{S\arabic{subsubsection}}
\subsubsection{Two-dimensional acousto-electric model for hexagonal piezoelectric materials}
\label{sec:2}
\noindent

 We use the surface impedance concept and logic developed by Ingebrigtsen \cite{ingebrigtsen1969surface} and adopt Gu\'eret's methlod for calculating the wave impedance \cite{gueret1971simple}. The following derivations are also based on the well-established theoretical framework from Farnell \textit{et al.} \cite{farnell1972elastic} and notations of Auld\cite{auld1973acoustic}. The model aims to describe the acousto-electric effect when carriers (here taken to be electrons) are introduced near the surface, forming a thin perturbed conducting layer on a hexagonal piezoelectric material (such as aluminum nitride). From Cristoffle's equations and Maxwell's equations, the coupled equations governing wave propagation in piezoelectric systems are
\begin{equation}
        \rho\frac{\partial^2\mathbf{u}}{\partial t^2}-\nabla\cdot \mathbf{c}^E:\nabla_s\mathbf{u}-\nabla \cdot (\mathbf{e}\cdot \nabla\phi)=0,
        \label{eq:S30}
\end{equation}
and
\begin{equation}
           \mathbf{e}:\nabla_s \mathbf{u} -\mathbf{\epsilon^S} \nabla\phi=0
           \label{eq:S31},
\end{equation}
where $\rho$ is the density of the piezoelectric material; $\mathbf{c}^E$ is the stiffness matrix at constant electric field; $\mathbf{u}$ is the particle-displacement vector; $\mathbf{e}$ is the piezoelectric matrix; $\mathbf{\epsilon^S}$ is the dielectric permittivity at constant strain; $\phi$ is the potential. The operators are defined as
\begin{equation}
    \nabla\cdot=
    \begin{bmatrix}
        \frac{\partial}{\partial x_1} &0&0&0&\frac{\partial}{\partial x_3}&\frac{\partial}{\partial x_2}\\
        0 &\frac{\partial}{\partial x_2}&0&\frac{\partial}{\partial x_3}&0&\frac{\partial}{\partial x_1}\\ 
        0 &0&\frac{\partial}{\partial x_3}&\frac{\partial}{\partial x_2}&\frac{\partial}{\partial x_1}&0
    \end{bmatrix},
    \label{eq:S32}
\end{equation}
\begin{equation}
    \nabla_s=
    \begin{bmatrix}
        \frac{\partial}{\partial x_1}&0&0\\
        0&\frac{\partial}{\partial x_2}&0\\
        0&0&\frac{\partial}{\partial x_3}\\
        0&\frac{\partial}{\partial x_3}&\frac{\partial}{\partial x_2}\\
        \frac{\partial}{\partial x_3}&0&\frac{\partial}{\partial x_1}\\
        \frac{\partial}{\partial x_2}&\frac{\partial}{\partial x_1}&0
    \end{bmatrix},
    \label{eq:S33}
\end{equation}
and
\begin{equation}
    \nabla=
    \begin{bmatrix}
        \frac{\partial}{\partial x_1}\\
        \frac{\partial}{\partial x_2}\\
        \frac{\partial}{\partial x_3}\\
    \end{bmatrix},
    \label{eq:S34}
\end{equation}
In the above equations, the quasi-static approximation has assumed
\begin{equation}
    \mathbf{E}=-\nabla\phi
    \label{eq:S35},
\end{equation}
where $\mathbf{E}$ is the electric field vector. The solutions of interests are straight-crested propagating waves which propagate along $x_1$ direction and have profiles along $x_3$ direction without displacement components along $x_2$ direction (same configuration with figure 1 of Farnell \cite{farnell1972elastic}).
\begin{equation}
    u_j=\alpha_j \exp(ikbx_3)\exp(ik(x_1-vt)), j=1,2,3,
    \label{eq:S36}
\end{equation}
and
\begin{equation}
    \phi=\alpha_4\exp(ikbx_3)\exp(ik(x_1-vt)),
    \label{eq:S37}
\end{equation}
where $b$ is the complex number that is defined by the coupled equations of Cristofle's equations and Maxwell's equations; $k$ is the wavenumber of the acoustic wave; $v$ is the velocity of the acoustic wave; $\alpha_j$ is the amplitude of the acousitc displacement or electric potential and $j=1,2,3$ corresponds to the direction $x_1$, $x_2$ or $x_3$. Since the above forms should satisfy the coupling equations, 4 equations about $\alpha_1$, $\alpha_2$, $\alpha_3$, $\alpha_4$ are governed by
\begin{equation}
    \begin{bmatrix}
        \Gamma_{11}-\rho v^2&\Gamma_{12} &\Gamma_{13} &\Gamma_{14}\\
        \Gamma_{12}&\Gamma_{22}-\rho v^2 &\Gamma_{23} &\Gamma_{24}\\
        \Gamma_{13}&\Gamma_{23} &\Gamma_{33}-\rho v^2 &\Gamma_{34}\\
        \Gamma_{14}&\Gamma_{24}&\Gamma_{34}&\Gamma_{44}
    \end{bmatrix}
    \begin{bmatrix}
        \alpha_1\\
        \alpha_2\\
        \alpha_3\\
        \alpha_4
    \end{bmatrix}=0,
    \label{eq:S38}
\end{equation}
where
\begin{equation}
\begin{bmatrix}
    \Gamma_{11}\\
    \Gamma_{22}\\
    \Gamma_{33}\\
    \Gamma_{12}\\
    \Gamma_{13}\\
    \Gamma_{23}\\
    \Gamma_{14}\\
    \Gamma_{24}\\
    \Gamma_{34}\\
    \Gamma_{44}\\
\end{bmatrix}=
\begin{bmatrix}
    c_{55}b^2+2c_{15}b+c_{11}\\
    c_{44}b^2+2c_{46}b+c_{66}\\
    c_{33}b^2+2c_{35}b+c_{55}\\
    c_{45}b^2+(c_{14}+c_{56})b+c_{16}\\
    c_{35}b^2+(c_{13}+c_{55})b+c_{15}\\
    c_{34}b^2+(c_{36}+c_{45})b+c_{56}\\
    e_{35}b^2+(e_{15}+e_{31})b+e_{11}\\
    e_{34}b^2+(e_{14}+e_{36})b+e_{16}\\
    e_{33}b^2+(e_{13}+e_{35})b+e_{15}\\
    -(\epsilon_{33}b^2+2\epsilon_{13}b+\epsilon_{11})

\end{bmatrix}.
\label{eq:S39}
\end{equation}
For hexagonal piezoelectric materials such as aluminum nitride, the stiffness matrix is 
\begin{equation}
    [c]=
    \begin{bmatrix}
        c_{11}& c_{12} & c_{13}& 0&0&0\\
        c_{12}& c_{11}&c_{13}&0&0&0\\
        c_{13}&c_{13}&c_{33}&0&0&0\\
        0&0&0&c_{44}&0&0\\
        0&0&0&0&c_{44}&0\\
        0&0&0&0&0&c_{66}
    \end{bmatrix}.
    \label{eq:S40}
\end{equation}
The equations become
\begin{equation}
    \begin{bmatrix}
        \Gamma_{11}-\rho v^2 &0&\Gamma_{13} &\Gamma_{14}\\
        0&\Gamma_{22}-\rho v^2&0&0\\
        \Gamma_{13}&0&\Gamma_{33}-\rho v^2 & \Gamma_{34}\\
        \Gamma_{14} &0&\Gamma_{34} &\Gamma_{44} 
    \end{bmatrix}
    \begin{bmatrix}
        \alpha_1\\
        \alpha_2\\
        \alpha_3\\
        \alpha_4
    \end{bmatrix}=0.
    \label{eq:S41}
\end{equation}
Therefore, the sagittal motion decouples from the transverse motion. Since only the sagittal waves are considered, the above equation becomes
\begin{equation}
    \begin{bmatrix}
        \Gamma_{11}-\rho v^2 &\Gamma_{13} &\Gamma_{14}\\
        \Gamma_{13}&\Gamma_{33}-\rho v^2 & \Gamma_{34}\\
        \Gamma_{14} &\Gamma_{34} &\Gamma_{44} 
    \end{bmatrix}
    \begin{bmatrix}
        \alpha_1\\
        \alpha_3\\
        \alpha_4
    \end{bmatrix}=0.
    \label{eq:S42}
\end{equation}
For nontrivial solutions of the field amplitude, the determinant must vanish and it yields
\begin{equation}
    \left|\Gamma_{rs}-\delta'_{rs}\rho v^2\right|=0, r,s=1,3,4, \delta'_{44}=0,
    \label{eq:S43}
\end{equation}
which is a sixth-order equation of the parameter $b$. For surface waves, the energy must decay in the bulk. Therefore, only three partial waves are allowed in the hexagonal piezoelectric material and the general solution is given by
\begin{equation}
    {u}_j=\sum_3C_n\alpha_j^{(n)} \exp(ikb^{(n)}x_3)\exp(ik(x_1-vt)),
    \label{eq:S44}
\end{equation}
and 
\begin{equation}
    {\phi}=\sum_3C_n\alpha_4^{(n)}\exp(ikb^{(n)}x_3)\exp(ik(x_1-vt))
    \label{eq:S45}.
\end{equation}
The specific form of the general solution is determined by boundary conditions. Since the conducting layer on top of the piezoelectric material is sufficiently thin, the mechanical boundary condition can still be approximated as that of a free top surface at $x_3=0$. The mechanical boundary condition indicates
\begin{equation}
    \begin{bmatrix}
        T_{11}&T_{12}&T_{13}\\
        T_{21}&T_{22}&T_{23}\\
        T_{31}&T_{32}&T_{33}\\
    \end{bmatrix}
    \begin{bmatrix}
        n_1\\
        n_2\\
        n_3\\
    \end{bmatrix}=
        \begin{bmatrix}
       0\\
       0\\
       0
    \end{bmatrix},
    \label{eq:S46}
\end{equation}
with
\begin{equation}
    \begin{bmatrix}
        n_1\\
        n_2\\
        n_3
    \end{bmatrix}=
    \begin{bmatrix}
        0\\
        0\\
        1
    \end{bmatrix}.
    \label{eq:S47}
\end{equation}
In Voigt notation, the mechanical boundary condition gives
\begin{equation}
    T_3=T_4=T_5=0.
    \label{eq:S48}
\end{equation}
Since
\begin{equation}
    \mathbf{S}=\frac{1}{i\omega}\nabla_S\mathbf{v},\mathbf{v}=\frac{\partial \mathbf{u}}{\partial t}=i\omega \mathbf{u},
    \label{eq:S49}
\end{equation}
where $\mathbf{S}$ is the strain and $\mathbf{v}$ is the particle velocity, the piezoelectric coupling can be written in terms of the particle displacement and the potential
\begin{equation}
     \mathbf{T}=c:\nabla_S\mathbf{u}+e\cdot \nabla\Phi,
     \label{eq:S50}
\end{equation}
and
\begin{equation}
    \mathbf{D}=-\epsilon\cdot \nabla\Phi+e:\nabla_S\mathbf{u}.
    \label{eq:S51}
\end{equation}
For hexagonal materials, the transverse motion is decoupled from the sagittal motion, $T_4=0$ automatically holds. We only need to take into account $T_3=0$ and $T_5=0$. $T_3=0$ yields
\begin{equation}
    c_{13}\frac{\partial u_1}{\partial x_1}+c_{33}\frac{\partial u_3}{\partial x_3}+e_{33}\frac{\partial \phi}{\partial x_3}=0,
    \label{eq:S52}
\end{equation}
which gives
\begin{equation}
    \sum_{3}C_n(c_{13}\alpha_1^{(n)}+c_{33}b^{(n)}\alpha_3^{(n)}+e_{33}b^{(n)}\alpha_4^{(n)})=0.
    \label{eq:S53}
\end{equation}
$T_5=0$ yields
\begin{equation}
    c_{44}\frac{\partial u_1}{\partial x_3}+c_{44}\frac{\partial u_3}{\partial x_1}+e_{15}\frac{\partial \phi}{\partial x_1}=0,
    \label{eq:S54}
\end{equation}
which gives
\begin{equation}
\sum_{3}C_n(c_{44}b^{(n)}\alpha_1^{(n)}+c_{44}\alpha_3^{(n)}+e_{15}\alpha_4^{(n)})=0.
    \label{eq:S55}
\end{equation}
The electric boundary condition requires the continuity of the vertical electric displacement at the free surface
\begin{equation}
   D_{3s}=-\varepsilon_{33}\frac{\partial \phi}{\partial x_3}+e_{31}\frac{\partial u_1}{\partial x_1}+e_{33}\frac{\partial u_3}{\partial x_3},
   \label{eq:S56}
\end{equation}
which gives
\begin{equation}
    D_{3s}=\sum_{3}C_nD^{(n)}_3\exp(ik(x_1-vt)),
    \label{eq:S57}
\end{equation}
where
\begin{equation}
    D^{(n)}_3=ik(e_{31}\alpha_1^{(n)}+e_{33}\alpha_3^{(n)}b^{(n)}-\varepsilon_{33}\alpha_4^{(n)}b^{(n)}).
    \label{eq:S58}
\end{equation}
Since 
\begin{equation}
    {\phi}=\sum_nC_n\alpha_4^{(n)}\exp(ikb^{(n)}x_3)\exp(ik(x_1-vt)),
    \label{eq:S59}
\end{equation}
at the surface ($x_3=0$)
\begin{equation}
    \phi_s=\sum_nC_n\alpha_4^{(n)}\exp(ik(x_1-vt)).
    \label{eq:S60}
\end{equation}
Note that mechanical boundary conditions give two independent equations for $C_1$, $C_2$ and $C_3$, the ratio between the potential and the electric displacement at the top surface is a constant
\begin{equation}
    \frac{\phi_s}{D_{3s}}=\frac{\sum_3C_n\alpha_4^{(n)}\exp(ik(x_1-vt))}{\sum_3C_nD_3^{(n)}\exp(ik(x_1-vt))}=\textit{const}.
    \label{eq:S61}
\end{equation}
Since we use the convention of $\exp(ikx-i\omega t)$ instead of $\exp(i\omega t-ikx)$ from the Ingebrigtsen's result, at steady states
\begin{equation}
    \phi_s=\frac{i}{k}E_1, D_{3s}=\frac{-k}{\omega}H_2.
    \label{eq:S62}
\end{equation}
The TM-wave surface impedance can be defined as
\begin{equation}
    Z_p=\frac{E_1}{H_2}=\frac{\omega i}{v^2}\frac{\phi_s}{D_{3s}}=(Z_{KA}(k^*))^*.
    \label{eq:S63}
\end{equation}
where $Z_{KA}$ is the impedance defined by Ingebrigtsen that gives
\begin{equation}
    Z_{KA}(k)=\frac{-i}{v\varepsilon}\frac{v_0-v}{v_\infty-v}=\frac{i}{v\varepsilon}\frac{k_0-k}{k_\infty-k}\frac{k_\infty}{k_0}\approx\frac{i}{v\varepsilon}\frac{k_0-k}{k_\infty-k},
    \label{eq:S64}
\end{equation}
Therefore,
\begin{equation}
    Z_p=\frac{-i}{v\varepsilon}\frac{k_0-k}{k_\infty-k},
    \label{eq:s65}
\end{equation}
where $k_0$ is the short-circuit wave number and $k_\infty$ is the open-circuit wave number. In carrier dynamics, The sheet impedance is governed by the drift-diffusion equation.
When there is a very thin conducting layer that the current density is approximately uniform inside the layer, the Amp\`ere's law indicates
\begin{equation}
    \mathbf{e}_3\times (\mathbf{H}_\textit{up}-\mathbf{H}_\textit{down})=J_s\mathbf{e}_1, J_s=Jh,
    \label{eq:S66}
\end{equation}
where $J_s$ is the surface current density; $J$ is the current density; $h$ is the thin layer thickness. The magnetic field is represented by
\begin{equation}
    \mathbf{H}_\textit{up/down}=H_\textit{tup/tdown}\mathbf{e}_2, \mathbf{e}_3\times \mathbf{e}_2=-\mathbf{e}_1,
    \label{eq:S67}
\end{equation}
where $H_\textit{tup}$ and $H_\textit{tdown}$ are values of tangential components at the upper interface between the conducting layer and air and at the lower interface between the conducting layer and the piezoelectric material surface.
The conducting layer provides the sheet admittance $Y_s$
\begin{equation}
    J_s\mathbf{e}_1=Y_sE_1\mathbf{e}_1,
    \label{eq:S68}
\end{equation}
which gives
\begin{equation}
    Y_s=\frac{J_s}{E_1}=\frac{H_\textit{tdown}-H_\textit{tup}}{E_1}.
    \label{eq:S69}
\end{equation}
Under the quasistatic approximation, the tangential magnetic field on the air side is negligible
\begin{equation}
    H_{tup} \approx 0.
    \label{eq:S70}
\end{equation}
The sheet is thin enough that the mass loading is neglected to preserve the free mechanical boundary condition, the sheet impedance becomes
\begin{equation}
    Y_s \approx \frac{H_\textit{tdown}}{E_1}= \frac{1}{Z_p}=(\frac{-i}{v\varepsilon}\frac{k_0-k}{k_\infty-k})^{-1}.
    \label{eq:S71}
\end{equation}
To calculate the sheet admittance from the carrier dynamics, the current density for electrons can be written as
\begin{equation}
    J=-q(n+fn_s)\mu (E_1+E_{DC})+\frac{\mu}{\beta}f\frac{\partial n_s}{\partial x_1},
    \label{eq:S72}
\end{equation}
where $E_1$ is the small traveling electric field associated with the acoustic field from piezoelectricity, $x_1$ is the propagation direction of the acoustic wave and the current density. All other notations are the same with those in one-dimensional acousto-electric model. We can expand it and neglect the higher-order term ($E_1n_s$)
\begin{equation}
        J=-qn\mu E_1-qn\mu E_{DC}-q\mu fn_sE_{DC}+\frac{\mu f}{\beta}\frac{\partial n_s}{\partial x_1}.
        \label{eq:S73}
\end{equation}
Neglect the trivial DC term ~$-qn\mu E_{DC}$. At steady states
\begin{equation}
        J=-qn\mu E_1-q\mu fn_sE_{DC}+\frac{\mu f}{\beta}ikn_s.
        \label{eq:S74}
\end{equation}
The continuity of the current density is given by
\begin{equation}
    \frac{\partial J}{\partial x_1}=-q\frac{\partial n_s}{\partial t},
    \label{eq:S75}
\end{equation}
and at steady states
\begin{equation}
    J=\frac{\omega}{k} qn_s.
    \label{eq:S76}
\end{equation}
Eliminate the current density term to get the expression of $n_s$ in terms of other parameters
\begin{equation}
    qn_s=\frac{kqn\mu E_1}{-\omega -k\mu fE_{DC}+\frac{\mu f}{\beta q}ik^2}.
    \label{eq:S77}
\end{equation}
From Gauss's law
\begin{equation}
    \nabla\cdot \mathbf{D}=qn_s,
    \label{eq:S78}
\end{equation}
and apply it to the sheet
\begin{equation}
    \int_{\partial V}\mathbf{D}(x_1,x_3)\cdot\mathbf{n}dS=\int_Vqn_sdV,
    \label{eq:S79}
\end{equation}
where the volume is selected as part of the sheet that is centered at $x_1=x$ and $x_3=h/2$. It has $h$ thickness from $x_3=0$ to $x_3=h$, $\delta$ length from $x_1=x_0-\delta/2$ to $x_1=x_0+\delta/2$ in the sagittal plane and $\Delta$ along the transverse direction. The above equation becomes
\begin{equation}
    (D_3(x,h^-)-D_3(x,0^+))\delta\Delta+h\Delta(D_1(x+\delta/2,h/2)-D_1(x-\delta/2,h/2))=qn_sh\delta \Delta,
    \label{eq:S80}
\end{equation}
and it can be simplified as
\begin{equation}
    \frac{ D_3(x,h^-)-D_3(x,0^+)}{h}+\frac{D_1(x+\delta/2,h/2)-D_1(x-\delta/2,h/2)}{\delta}=qn_s,
    \label{eq:S81}
\end{equation}
where $D_3$ and $D_1$ are components of $\mathbf{D}$ in the sagittal plane. Since $\delta$ is small, it is approximately the same with the derivative form
\begin{equation}
    D_3(x,h^-)-D_3(x,0^+)+h\frac{\partial D_1}{\partial x_1}=qn_sh.
    \label{eq:S82}
\end{equation}
From Maxwell's equation
\begin{equation}
    \nabla\times\mathbf{H}=\frac{\partial \mathbf{D}}{\partial t}+J,
    \label{eq:S83}
\end{equation}
We have
\begin{equation}
    kH_2=-\omega D_3,
    \label{eq:S84}
\end{equation}
which gives
\begin{equation}
    H_2(x,h^-)-H_2(x,0^+)=\frac{-\omega}{k}( D_3(x,h^-)-D_3(x,0^+)).
    \label{eq:S85}
\end{equation}
Due to the continuity of $H_2,D_3$ at the boundary of the sheet, we can rewrite the above equation as
\begin{equation}
    H_2(x,h)-H_2(x,0)=\frac{-\omega}{k}( D_3(x,h)-D_3(x,0)).
    \label{eq:S86}
\end{equation}
Since the sheet admittance at steady states is given by
\begin{equation}
    Y_s=\frac{H_\textit{tdown}-H_\textit{tup}}{E_1}=\frac{\omega( D_3(x,h)-D_3(x,0))}{kE_1}=\frac{\omega(qn_sh-ihkD_1)}{kE_1}.
    \label{eq:S87}
\end{equation}
Substitute the $n_s$ term by the drift-diffusion equation
\begin{equation}
    Y_s=\frac{\omega(\frac{kqn\mu E_1}{-\omega -k\mu fE_{DC}+\frac{\mu f}{\beta q}ik^2}h-ihkD_1)}{kE_1}.
    \label{eq:S88}
\end{equation}
If we neglect the piezoelectric coupling correction
\begin{equation}
    D_1=\varepsilon E_1,
    \label{eq:S89}
\end{equation}
the sheet impedance becomes
\begin{equation}
    Z_s=\frac{k}{\omega(\frac{kqn\mu }{-\omega -k\mu fE_{DC}+\frac{\mu f}{\beta q}ik^2}h-ihk\varepsilon)},
    \label{eq:S90}
\end{equation}
which is
\begin{equation}
    Z_s=\frac{-ki}{\omega\varepsilon}(kh(-1+\frac{iqn\mu/\varepsilon}{\omega +k\mu fE_{DC}-\frac{\mu f}{\beta q}ik^2}))^{-1}.
    \label{eq:S91}
\end{equation}
To simplify the notations, we define
\begin{equation}
    \omega_c=\frac{n|q|\mu}{\varepsilon},\omega_D=\frac{|q|\beta}{f\mu}v^2, v=\frac{\omega}{k},
    \label{eq:S92}
\end{equation}
and the sheet impedance is
\begin{equation}
    Z_s=\frac{-i}{v\varepsilon}(kh(-1+\frac{-i\omega_c/\omega }{1 +\mu fE_{DC}/v+i\frac{\omega}{\omega_D}}))^{-1}. 
    \label{eq:S93}
\end{equation}
Since the sheet impedance is
\begin{equation}
    Z_s=-\frac{i}{v\varepsilon}\frac{k_0-k}{k_\infty-k}
    \label{eq:S94},
\end{equation}
where $k$ here is the complex wave number, we have
\begin{equation}
    \frac{k_0-k}{k_\infty-k}=\frac{1}{(-1+\frac{-i\omega_c/\omega}{1+\mu fE_{DC}/v+i\frac{\omega}{\omega_D}})kh},
    \label{eq:S95}
\end{equation}
which is
\begin{equation}
    k-k_{\infty}=-kh(k_0-k)(-1+\frac{-i\omega_c/\omega}{1+\mu fE_{DC}/v+i\frac{\omega}{\omega_D}}),
    \label{eq:S96}
\end{equation}
In weak piezoelectric regime
\begin{equation}
    k_0=k_{\infty}(1+\frac{\Delta v_\infty}{v})=k_\infty(1+\frac{K^2}{2})
    \label{eq:S97}
\end{equation}
where $k_0$ is the wave number for insulating sheet. ~$k_\infty$ is the wave number for conducting sheet. Define a complex number
\begin{equation}
    X_0=1+\frac{i\omega_c/\omega}{1+\mu fE_{DC}/v+i\frac{\omega}{\omega_D}}.
    \label{eq:S98}
\end{equation}
and simplify the terms
\begin{equation}
    \frac{1}{X_0}=\frac{(1+\frac{\mu f E_{DC}}{v})^2-i\frac{\omega_c}{\omega}(1+\frac{\mu fE_{DC}}{v})+\frac{\omega}{\omega_D}(\frac{\omega}{\omega_D}+\frac{\omega_c}{\omega})}{(1+\frac{\mu fE_{DC}}{v})^2+(\frac{\omega}{\omega_D}+\frac{\omega_c}{\omega})^2}=R+Ii.
    \label{eq:S99}
\end{equation}
The equation becomes
\begin{equation}
    \frac{k}{X_0}+hk^2=\frac{k_\infty}{X_0}+k_\infty(1+\frac{K^2}{2})hk
    \label{eq:S100}.
\end{equation}
To find the attenuation of acoustic waves, we substitute $k$ with $k-i\alpha$, where $\alpha$ is defined as the same gain coefficient (spatial growth rate) with that in one-dimensional model section. Under small gain region
\begin{equation}
    (R+iI)(k-i\alpha)+h(k^2-2i\alpha k)=k_\infty(R+iI)+k_\infty(1+\frac{K^2}{2})h(k-i\alpha).
    \label{eq:S101}
\end{equation}
Apparently, when $h\to 0, k \to k_\infty, \alpha \to 0$. To find the attenuation, the imaginary part should be equal, yielding
\begin{equation}
    \alpha=\frac{I(k-k_\infty)}{R+\gamma},
    \label{eq:S102}
\end{equation}
where
\begin{equation}
    \gamma=2kh-k_\infty h(1+\frac{K^2}{2})<k_0h.
    \label{eq:S103}
\end{equation}
The gain coefficient becomes
\begin{equation}
    \alpha=-\frac{\frac{\omega_c}{\omega}(1+\frac{\mu fE_{DC}}{v})(k-k_\infty)}{(1+\gamma)(1+\frac{\mu fE_{DC}}{v})^2+\frac{\omega}{\omega_D}(\frac{\omega}{\omega_D}+\frac{\omega_c}{\omega})+\gamma(\frac{\omega}{\omega_D}+\frac{\omega_c}{\omega})^2}.
    \label{eq:S104}
\end{equation}
Since the origin of the above derivations (sheet impedance from the carrier dynamics is equal to the surface impedance from acoustic dynamics) assumes the sheet is thin enough that the mass loading to the bulk piezoelectric material is neglected, we have
\begin{equation}
    k_0h\ll1.
    \label{eq:S105}
\end{equation}
In the week coupling regime, the wave vector is expanded as
\begin{equation}
    k(h)=k(0)+k'(0)h=k_\infty+k'(0) h,
    \label{eq:S106}
\end{equation}
where $k'(0)$ is the slope of the wave-number-vs-sheet-thickness curve at zero sheet thickness. When the thickness is zero, no sheet exists and the wave number corresponds to the open-circuit case. Notice that
\begin{equation}
    k-k_\infty<k_0-k_\infty=\frac{K^2}{2}k_\infty,
    \label{eq:S107}
\end{equation}
and for electrons
\begin{equation}
    v_e=-\mu E_{DC},
    \label{eq:S108}
\end{equation}
the gain coefficient is
\begin{equation}
    \alpha=\frac{-\frac{\omega_c}{\omega}(1-\frac{fv_e}{v})k'(0) h}{(1-\frac{fv_e}{v})^2+\frac{\omega}{\omega_D}(\frac{\omega}{\omega_D}+\frac{\omega_c}{\omega})+\gamma(\frac{\omega}{\omega_D}+\frac{\omega_c}{\omega})^2},k'(0)h<\frac{K^2}{2}k_\infty.
    \label{eq:S109}
\end{equation}
It has a similar form compared with that in the one-dimensional model. 
\setcounter{secnumdepth}{3}
\renewcommand{\thesubsubsection}{S\arabic{subsubsection}}
\subsubsection{Buried sheet in two-dimensional acousto-electric model}
\label{sec:3}
When the sheet is buried inside the piezoelectric material at the depth $x_3=-H$, the sheet admittance is given by
\begin{equation}
    Y_s=\frac{H_\textit{tdown}-H_\textit{tup}}{E_1}=\frac{1}{Z_\textit{down}}-\frac{1}{Z_\textit{up}},
    \label{eq:S110}
\end{equation}
where $Z_\textit{up}$ and $Z_\textit{down}$ are the impedances seen from the top and bottom surfaces of the sheet toward the free piezoelectric surface and the bulk piezoelectric medium, respectively. In general, the unperturbed impedance satisfies
\begin{equation}
    Z_d(x_3)=\frac{\omega}{iv^2}\frac{\phi_s(x_3)}{D_{3s}(x_3)}=\frac{\omega}{iv^2}\frac{\sum_3C_n\alpha_4^{(n)}\exp(ikb^{(n)}x_3)\exp(ik(x_1-vt))}{\sum_3C_nD_3^{(n)}\exp(ikb^{(n)}x_3)\exp(ik(x_1-vt))}.
    \label{eq:S111}
\end{equation}
Since $Z_d(x_3)$ is the impedance at $x_3$ seen into the bulk substrate in the absence of the sheet, the downward impedance is
\begin{equation}
    Z_\textit{down} = Z_d(-H-h/2),
    \label{eq:S112}
\end{equation}
provided that the sheet produces a small perturbation. Because the impedance depends on the mode-dependent coefficents $b^{(n)}$, $Z_\textit{up}$ is generally different for different acoustic modes. Since $Z_\textit{up}$ is the impedance looking from $x_3$ into the air half-space, it is finite when $x_3\neq 0$. Therefore, a buried sheet can yield the mode-dependent acousto-electric effect.
\setcounter{secnumdepth}{3}
\renewcommand{\thesubsubsection}{S\arabic{subsubsection}}
\subsubsection{Material parameters for AlN-SOI simulation}
\label{sec:4}
The aluminum nitride (AlN) polycrystalline thin film deposited on the SOI platform is isotropic in-plane and the c-axis of the aluminum nitride is aligned with the out-of-plane direction. In COMSOL Multiphysics, the elastic tensor and the piezoelectric tensor correspond the single crystal AlN. It is important to understand the difference between these two cases to model our device correctly in the simulation software. B. A. Auld\cite{auld1973acoustic} refers to the Bond transformation properties \cite{bond1943mathematics} for this type of problem. The elasticity matrix of the AlN single crystal is given by
\begin{equation}
    [c]=
    \begin{bmatrix}
        c_{11}& c_{12} & c_{13}& 0&0&0\\
        c_{12}& c_{11}&c_{13}&0&0&0\\
        c_{13}&c_{13}&c_{33}&0&0&0\\
        0&0&0&c_{44}&0&0\\
        0&0&0&0&c_{44}&0\\
        0&0&0&0&0&c_{66}
    \end{bmatrix},
    \label{eq:S113}
\end{equation}
where we use the same coordinate system as that in the two-dimensional acousto-electric model. For each grain in the polycrystalline thin film, the in-plane rotation angle is $\delta$ and the rotation matrix is
\begin{equation}
    [a]=
    \begin{bmatrix}
        \cos(\delta) &\sin(\delta) &0\\
        -\sin(\delta) & \cos(\delta) & 0\\
        0&0&1
    \end{bmatrix}.
    \label{eq:S114}
\end{equation}
Before rotation, the piezoelectric coupling requires
\begin{equation}
     [T]=-[e][E]+[c][S],
     \label{eq:S115}
\end{equation}
After rotation
\begin{equation}
    [E']=[a][E],[T']=[M][T],[S']=[M][S],
    \label{eq:S116}
\end{equation}
with the new piezoelectric coupling
\begin{equation}
    [T']=-[e'][E']+[c'][S'],
    \label{eq:S117}
\end{equation}
which becomes
\begin{equation}
    [M][T]=-[e'][a][E]+[c'][M][S].
    \label{eq:S118}
\end{equation}
By comparing it with the piezoelectric coupling before rotation
\begin{equation}
    [e']=[M][e][a^{-1}],[c']=[M][c][M^{-1}],
    \label{eq:S119}
\end{equation}
the transformation matrix is attached to elements $a_{ij}$ in the rotation matrix $[a]$
\begin{equation}
    [M]=\begin{bmatrix}
        a_{11}^2&a_{12}^2&a_{13}^2&2a_{12}a_{13}& 2a_{13}a_{11}&2a_{11}a_{12}\\
        a_{21}^2&a_{22}^2&a_{23}^2&2a_{22}a_{23}& 2a_{23}a_{21}&2a_{21}a_{22}\\
        a_{31}^2&a_{32}^2&a_{33}^2&2a_{32}a_{33}& 2a_{33}a_{31}&2a_{31}a_{32}\\
        a_{21}a_{31}&a_{22}a_{32}&a_{23}a_{33}&a_{22}a_{33}+a_{23}a_{32}&a_{21}a_{33}+a_{23}a_{31}&a_{22}a_{31}+a_{21}a_{32}\\
        a_{31}a_{11}&a_{32}a_{12}&a_{33}a_{13}&a_{12}a_{33}+a_{13}a_{32}&a_{13}a_{31}+a_{11}a_{33}&a_{11}a_{32}+a_{12}a_{31}\\
        a_{11}a_{21}&a_{12}a_{22}&a_{13}a_{23}&a_{12}a_{23}+a_{13}a_{22}&a_{13}a_{21}+a_{11}a_{23}&a_{11}a_{22}+a_{12}a_{21}
    \end{bmatrix}.
    \label{eq:S120}
\end{equation}
When  the rotation is around c-axis ($x_3$ direction)
\begin{equation}
    [M]=
    \begin{bmatrix}
        \cos^2(\delta)&\sin^2{\delta} &0&0&0&\sin(2\delta)\\
        \sin^2(\delta)&\cos^2{\delta} &0&0&0&-\sin(2\delta)\\
        0&0&1&0&0&0\\
        0&0&0&\cos(\delta) &-\sin(\delta)&0\\
        0&0&0&\sin(\delta)&\cos(\delta)&0\\
        -\sin(2\delta)/2 &\sin(2\delta)/2&0&0&0&\cos(2\delta)
    \end{bmatrix}.
    \label{eq:S121}
\end{equation}
Since
\begin{equation}
    [M^{-1}]=[M^{T}],
    \label{eq:S122}
\end{equation}
the elastic matrix becomes
\begin{equation}
    [c']=
    \begin{bmatrix}
        c_{n11}& c_{n12} & c_{13}& 0&0&c_{n16}\\
        c_{n12}& c_{n11}&c_{13}&0&0&-c_{n16}\\
        c_{13}&c_{13}&c_{33}&0&0&0\\
        0&0&0&c_{44}&0&0\\
        0&0&0&0&c_{44}&0\\
        c_{n16}&-c_{n16}&0&0&0&c_{n66}
    \end{bmatrix},
    \label{eq:S123}
\end{equation}
where
\begin{equation}
    \begin{bmatrix}
        c_{n11}\\
        c_{n12}\\
        c_{n16}\\
        c_{n66}\\
    \end{bmatrix}=
    \begin{bmatrix}
        c_{11}+\frac{1-\cos(4\delta)}{4}(2c_{66}-c_{11}+c_{12})\\
        c_{12}-\frac{1-\cos(4\delta)}{4}(2c_{66}-c_{11}+c_{12})\\
        (2c_{66}-c_{11}+c_{12})\frac{\sin(4\delta)}{4}\\
        \frac{c_{11}-c_{12}}{2}+(2c_{66}-c_{11}+c_{12})\frac{1+\cos(4\delta)}{4}
    \end{bmatrix}.
    \label{eq:S124}
\end{equation}
Since AlN belongs to the hexagonal system, it should satisfy
\begin{equation}
    c_{66}=\frac{1}{2}(c_{11}-c_{12}).
    \label{eq:S125}
\end{equation}
As a result, for any rotation angle $\delta$,
\begin{equation}
    [c']=[c].
    \label{eq:S126}
\end{equation}
The piezoelectric stress matrix in AlN is
\begin{equation}
    [e]=\begin{bmatrix}
        0&0&e_{31}\\
        0&0&e_{31}\\
        0&0&e_{33}\\
        0&e_{15}&0\\
        e_{15}&0&0\\
        0&0&0
        
    \end{bmatrix}.
    \label{eq:S127}
\end{equation}
Similarly, the piezoelectric matrix remains unchanged after rotation, yielding
\begin{equation}
    [e']=[e].
    \label{eq:S128}
\end{equation}
Therefore, it is reasonable to simulate the acoustic structure of the polycrystalline AlN film using single-crystal material parameters.
\setcounter{secnumdepth}{3}
\renewcommand{\thesubsubsection}{S\arabic{subsubsection}}
\subsubsection{S-parameter simulation of the AlN-SOI platform}
\label{sec:5}
\noindent
\begin{figure}[!b] 
	\centering
	\includegraphics[width=0.8\textwidth]{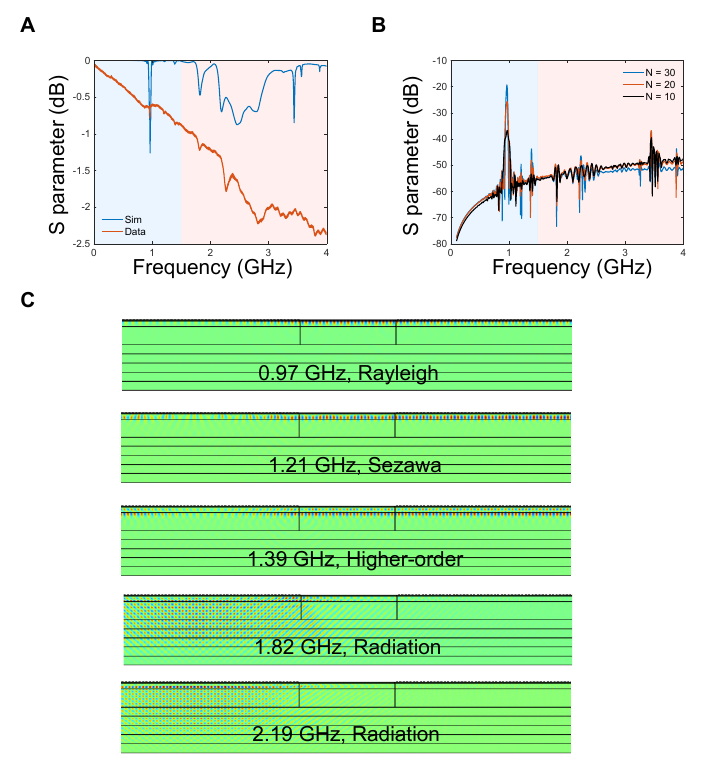} 
	\caption{\textbf{RF simulation of the AlN-SOI platform.}
		(A) $\mathrm{S_{11}}$ simulation and the experimental data of the device in the main text (Figure~\ref{fig:2}C). The shallow blue region corresponds to the SAW response while the shallow red region corresponds to the radiation response. (B) $\mathrm{S_{21}}$ simulation with different numbers of IDT finger pairs. (C) Mode profiles (vertical displacement) at different $\mathrm{S_{11}}$ dips. 
		}
	\label{fig:S2} 
\end{figure}
We use COMSOL Multiphysics to investigate acoustic properties below $4~\mathrm{GHz}$ of the AlN-SOI platform and the elastic properties of AlN are taken from Mehwish \textit{et al.} \cite{aslam2018fem}. The simulation geometry matches the device studied in the main text with minor adjustments. First, the length of the delay line is decreased to reduce the computation cost. Second,the AlN etch used for electrode deposition is neglected, as it primarily introduces additional scattering of the Rayleigh wave but does not substantially change the overall elastic properties of the structure. Third, to suppress the unwanted reflections, we apply manually defined absorbing boundary regions on the left, right, and bottom boundaries using a graded increase of the Rayleigh damping, rather than using a perfectly matched layer (PML). We adopt this approach because the PML boundary condition can introduce non-physical energy localization at the PML-domain interface for complex multilayer structures. In Fig.~\ref{fig:S2}(A), the simulated $\mathrm{S_{11}}$ agrees with the measured data, especially in the light blue region where the Rayleigh, Sezawa and higher-order mode are observed. In the light red region, the large dips correspond to bulk-wave radiation. The linear decrease with increasing frequency observed in the measured data, but absent in the simulation, is interpreted as an effective capacitance associated with the IDT transducer. The simulated $\mathrm{S_{21}}$ is shown in Fig.~\ref{fig:S2}(B). As the number of IDT finger pairs increases, constructive interference enhances the Sezawa and the higher-order mode, making them more significant. No bulk modes are observed in this plot because outgoing waves are absorbed at the boundaries of the simulation domain. In the experiment, however, imperfectly absorbing boundaries can convert radiated bulk waves into guided bulk modes. Mode profiles are shown in Fig.~\ref{fig:S2}(C). 
\setcounter{secnumdepth}{3}
\renewcommand{\thesubsubsection}{S\arabic{subsubsection}}
\subsubsection{Acousto-optic scattering model }
\label{sec:6}
In the interaction region, the acoustic wave time-perturbs the permittivity of the silicon ridge waveguide with the acoustic angular frequency $\Omega$ that yields
\begin{equation}
    \varepsilon(z,t)=\varepsilon+\delta\varepsilon e^{i(qz-\Omega t)}+\delta\varepsilon^*e^{-i(qz-\Omega t)}.
    \label{eq:S129}
\end{equation}
The optical field satisfies Maxwell's equation
\begin{equation}
    \nabla\times\nabla\times\mathbf{E}+\mu_0\frac{\partial^2}{\partial t^2}(\varepsilon(z,t)\mathbf{E})=0.
    \label{eq:S130}
\end{equation}
We expand the total electric field as the sum of a pump, a Stokes field, and an anti-Stokes field
\begin{equation}
    \mathbf{E}=(a_p(z)e^{i(\beta_p z-\omega_p t)}+a_s(z)e^{i(\beta_s z-\omega_s t)}+a_{as}(z)e^{i(\beta_{as} z-\omega_{as} t)})\mathbf{e_\perp}+c.c..
    \label{eq:S131}
\end{equation}
$a_p(z)$, $a_s(z)$ and $a_{as}(z)$ are the evolution of envelops for the pump, Stokes and anti-Stokes fields, respectively; $\mathbf{e}_\perp$ denotes the transverse mode profile and its polarization (TE) inside the waveguide; $c.c.$ are complex conjugate terms. From Maxwell's equation, the sideband frequencies satisfy
\begin{equation}
    \omega_s=\omega_p-\Omega,
    \label{eq:S132}
\end{equation}
and
\begin{equation}
    \omega_{as}=\omega_p+\Omega.
    \label{eq:S133}
\end{equation}
For compactness, we define
\begin{equation}
    E_p=a_p(z)e^{i(\beta_pz-\omega_p t)}, E_s=a_s(z)e^{i(\beta_sz-\omega_s t)}, E_{as}=a_{as}(z)e^{i(\beta_{as}z-\omega_{as} t)}.
    \label{eq:S134}
\end{equation}
The driving terms relating to $\delta\varepsilon$ from the pump, Stokes and anti-Stokes fields are
\begin{equation}
    \delta\varepsilon e^{i(qz-\Omega t)}E_p=\delta\varepsilon a_p(z)e^{i(\beta_p+q)z}e^{-i(\omega_p+\Omega)t},
    \label{eq:S135}
\end{equation}
\begin{equation}
    \delta\varepsilon e^{i(qz-\Omega t)}E_s=\delta\varepsilon a_s(z)e^{i(\beta_s+q)z}e^{-i(\omega_s+\Omega)t},
    \label{eq:S136}
\end{equation}
and
\begin{equation}
    \delta\varepsilon e^{i(qz-\Omega t)}E_{as}=\delta\varepsilon a_{as}(z)e^{i(\beta_{as}+q)z}e^{-i(\omega_{as}+\Omega)t},    
    \label{eq:S137}
\end{equation}
The driving term relating to $\delta\varepsilon^*$ from the pump, Stokes and anti-Stokes fields are
\begin{equation}
    \delta\varepsilon^*e^{-i(qz-\Omega t)}E_p=\delta\varepsilon^* a_p(z)e^{i(\beta_p-q)z}e^{-i(\omega_p-\Omega)t},
    \label{eq:S138}
\end{equation}
\begin{equation}
    \delta\varepsilon^* e^{-i(qz-\Omega t)}E_s=\delta\varepsilon^* a_s(z)e^{i(\beta_s-q)z}e^{-i(\omega_s-\Omega)t},
    \label{eq:S139}
\end{equation}
and
\begin{equation}
    \delta\varepsilon^* e^{-i(qz-\Omega t)}E_{as}=\delta\varepsilon^* a_{as}(z)e^{i(\beta_{as}-q)z}e^{-i(\omega_{as}-\Omega)t},    
    \label{eq:S140}
\end{equation}
Since the higher-order sidebands with angular frequency $\omega_{as}+\Omega$ and $\omega_s-\Omega$ are negligible in the experiment, we only consider the interaction among pump, Stokes and anti-Stokes fields. For the Stokes field, it satisfies
\begin{equation}
    \nabla\times\nabla\times(E_s\mathbf{e_\perp})+\mu_0\frac{\partial^2}{\partial t^2}(\varepsilon E_s\mathbf{e_\perp})=-\mu_0\frac{\partial^2}{\partial t^2}(\delta\varepsilon^*E_pe^{-i(qz-\Omega t)}\mathbf{e_\perp}),
    \label{eq:S141}
\end{equation}
which is
\begin{equation}
    \nabla\times\nabla\times(a_s(z)e^{i\beta_sz}\mathbf{e_\perp})-\mu_0\varepsilon\omega_s^2a_s(z)e^{i\beta_sz}\mathbf{e_\perp}=\mu_0\omega_s^2\delta\varepsilon^*a_p(z)e^{i(\beta_p-q)z}\mathbf{e_\perp}.
    \label{eq:S142}
\end{equation}
To solve this equation, we denote 
\begin{equation}
    \mathbf{F}_s=e^{i\beta_sz}\mathbf{e_\perp}=e^{i\beta_sz}e(x,y)\mathbf{y},
    \label{eq:S143}
\end{equation}
where $e(x,y)$ is the TE mode profile and $\mathbf{y}$ is the mode polarization direction. Since
\begin{equation}
    \nabla\times\nabla\times(a_s(z)\mathbf{F}_s)=\nabla\times(\nabla a_s(z)\times \mathbf{F}_s)+\nabla a_s(z)\times(\nabla\times\mathbf{F}_s)+a_s(z)\nabla\times\nabla\times\mathbf{F}_s,
    \label{eq:S144}
\end{equation}
the above equation becomes
\begin{equation}
    \nabla\times(\nabla a_s(z)\times \mathbf{F}_s)+\nabla a_s(z)\times(\nabla\times\mathbf{F}_s)=\mu_0\omega_s^2\delta\varepsilon^*a_p(z)e^{i(\beta_p-q)z}\mathbf{e_\perp}.
    \label{eq:S145}
\end{equation}
Note that the envelop is independent of the transverse mode profile, we have
\begin{equation}
    \nabla a_s(z)=\mathbf{z}\frac{d a_s(z)}{dz},
    \label{eq:S146}
\end{equation}
and the above equation can be further simplified as
\begin{equation}
    -(a_s''+2i\beta_sa_s')\mathbf{F}_s+a_s'\frac{\partial e(x,y)}{\partial x}e^{i\beta_s z}\mathbf{z}=\mu_0\omega_s^2\delta\varepsilon^*a_p(z)e^{i(\beta_p-q)z}\mathbf{e_\perp}.
    \label{eq:S147}
\end{equation}
The apparent inconsistency in the above equation arises from the simplifying assumption that the optical mode is purely transverse. However, this inconsistent longitudinal correction term can be neglected because the optical propagation constant is much larger than the characteristic transverse variation of the mode profile
\begin{equation}
    \beta_s \gg\frac{\partial e(x,y)}{\partial x}\frac{1}{e(x,y)}\approx\frac{1}{w}.
    \label{eq:S148}
\end{equation}
where $w$ is the characteristic width that confines the optical modes. From slow-varying envelope approximation, the second derivative can also be neglected, yielding
\begin{equation}
    -2i\beta_sa_s'\mathbf{F}_s=\mu_0\omega_s^2\delta\varepsilon^*a_p(z)e^{i(\beta_p-q)z}\mathbf{e_\perp},
    \label{eq:S149}    
\end{equation}
which gives
\begin{equation}
    \frac{da_s}{dz}=i\frac{\mu_0\omega_s^2\delta\varepsilon^*}{2\beta_s}a_pe^{i(\beta_p-\beta_s-q)z}.
    \label{eq:S150}
\end{equation}
For the anti-Stokes field, it satisfies 
\begin{equation}
     \nabla\times\nabla\times(E_{as}\mathbf{e_\perp})+\mu_0\frac{\partial^2}{\partial t^2}(\varepsilon E_{as}\mathbf{e_\perp})=-\mu_0\frac{\partial^2}{\partial t^2}(\delta\varepsilon E_pe^{i(qz-\Omega t)}\mathbf{e_\perp}),
     \label{eq:S151}
\end{equation}
which gives
\begin{equation}
    \frac{da_{as}}{dz}=i\frac{\mu_0\omega_{as}^2\delta\varepsilon}{2\beta_{as}}a_pe^{i(\beta_p+q-\beta_{as})z}.
    \label{eq:S152}
\end{equation}
In the experiment, the acoustic field propagates perpendicular to the waveguide, so the longitudinal acoustic wave number along the optical propagation direction is
\begin{equation}
    q=0.
    \label{eq:S153}
\end{equation}
The phase mismatch can then be estimated as (assume the optical group velocity is $1\times 10^8 ~\mathrm{m/s}$ and the interaction length $L$ is $300~\mathrm{\mu m}$)
\begin{equation}
    \Delta \beta_s L=(\beta_p-\beta_s)L= 0.027
    \label{eq:S154}
\end{equation}
We assume the pump field is undepleted, and the generated Stokes power is given by
\begin{equation}
    P_s(L)=|\kappa_s|^2P_pL^2\mathrm{sinc}^2(\frac{\Delta \beta_sL}{2}),
    \label{eq:S155}
\end{equation}
where the spatial coupling rate $\kappa_s$ is defined as
\begin{equation}
    |\kappa_s|=|\frac{\mu_0\omega_s^2\delta \varepsilon^*}{2\beta_s}|.
    \label{eq:S156}
\end{equation}
\begin{figure} 
	\centering
	\includegraphics[width=1\textwidth]{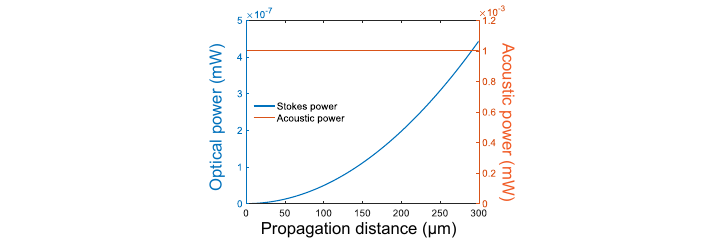} 
	\caption{\textbf{Simulated Brillouin sidebands under acoustic drive.}
		Pump power: $0~\mathrm{dBm}$; acoustic power: $-30~\mathrm{dBm}$. The Stokes and anti-Stokes power are identical. 
		}
	\label{fig:S3} 
\end{figure}
The anti-Stokes field has similar dynamics. To extract the permittivity change of the device under test in the main text, we use the above equations of motion together with the experimental parameters to establish the dynamics. In the experiment, the input optical power is $0~\mathrm{dBm}$, and the RF input power on the IDT is $0~\mathrm{dBm}$. Since the $S_{21}$ parameter is around $-60~\mathrm{dB}$ for the Sezawa mode, the acoustic power in the delay line is approximately $-30~\mathrm{dBm}$. The RF amplifier after heterodyne detection gives $20~\mathrm{dB}$ gain and the responsivity of the photodiode is $0.85~\mathrm{A/W}$ with $500~\mathrm{\Omega}$ transimpedance gain, yielding a measured Stokes-related signal power of $-60~\mathrm{dBm}$ at the vector network analyzer. Therefore, the beat note between the Stokes field and the local oscillator (LO) has the power of $5.3\times 10^{-5}~\mathrm{mW}$. Since the LO is $-7~\mathrm{dBm}$, the power of the Stokes component is $1.4\times 10^{-8}~\mathrm{mW}$. The loss from the Brillouin interaction region to the photodetector termination is assumed to be $-15~\mathrm{dB}$ ($-12~\mathrm{dB}$ from the grating coupler and $-3~\mathrm{dB}$ from the beam splitter). The generated Stokes power can then be back-calculated as $4.43\times 10^{-7}~\mathrm{mW}$. Moreover, the effective index is assumed to be $n_{eff}=2.4$ and this result gives the relative permittivity change $\delta\varepsilon_r=2.6\times 10^{-6}$.   Fig.~\ref{fig:S3} shows the corresponding dynamics in the interaction region that generate this level of Stokes power.
\setcounter{secnumdepth}{3}
\renewcommand{\thesubsubsection}{S\arabic{subsubsection}}
\subsubsection{The time-gated signal}
\label{sec:7}
\begin{figure}[!b] 
	\centering
	\includegraphics[width=1\textwidth]{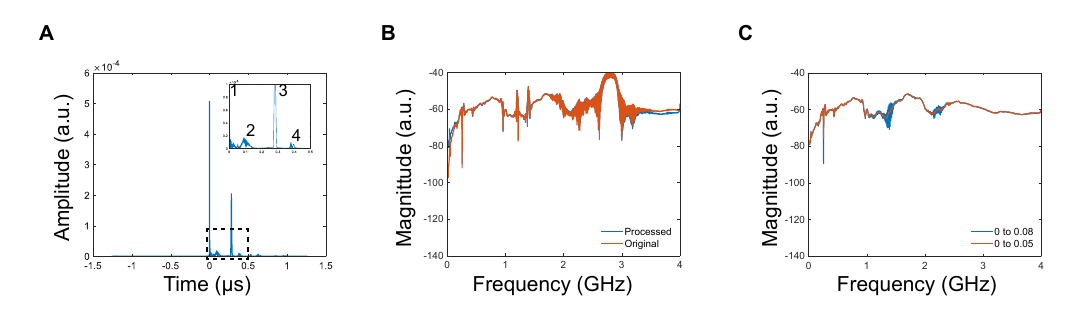} 
	\caption{\textbf{Time gated signals of the $S_{21}$ spectrum without external voltage.}
		(A) The time sequence constructed after inverse Fourier transform of the spectrum. (B) Original spectrum versus the spectrum after processing without applying any time-domain filtering (inverse Fourier transform, then Fourier transform back). (C) Time-gated spectra obtained by retaining only the $0\to 0.05~\mathrm{\mu s}$ or $0\to 0.08 ~\mathrm{\mu s}$ time window and removing (zeroing) the remaining portions of the time-domain signal.
		}
	\label{fig:S4} 
\end{figure}
In the AlN-SOI platform, because the signal from the surface acoustic wave (SAW) branch is fully buried within the strong electromagnetic feedthrough, a careful time-gating process is necessary to remove it without introducing artifacts into the signal. In Fig.~\ref{fig:S4}(A),the time-domain signal is obtained by applying an inverse Fourier transform to the spectrum. Four significant peaks, labeled from 1 through 4, correspond to (1) the instantaneous electromagnetic feedthrough, (2) surface acoustic wave (SAW) propagation, (3) bulk waves radiating from the interdigital transducer region (supported by the device’s bottom boundary condition), and (4) multiple reflected waves arising from triple-pass transmission. To evaluate how the frequency window from $20~\mathrm{MHz}$ to $4~\mathrm{GHz}$ affects the Fourier transform, Fig.~\ref{fig:S4}(B) compares the processed spectrum with the original spectrum. The strong overlap between the two spectra indicates that the windowing effect is negligible. In the main text, all time-gated signals exclude the time response before $0.05~\mathrm{\mu s}$. Fig.~\ref{fig:S4}(C) is used to examine frequency components after time gating. When only the early-time portion of the signal is retained (from $0~\mathrm{\mu s}$ to $0.05~\mathrm{\mu s}$), the resulting spectrum shows a smooth background curve without interference fringes, indicating that no SAW branch information is present within this time interval. In contrast, when the retained window is extended to $0.08~\mathrm{\mu s}$, fringes begin to appear. We further tested the recovered spectrum by removing the time window before $0.4~\mathrm{\mu s}$, $0.5~\mathrm{\mu s}$ and $0.6~\mathrm{\mu s}$. In all cases, the recovered SAW mode magnitudes remain the same, showing that the SAW modes can be reliably recovered even when they are buried beneath the strong electromagnetic feedthrough. 
\setcounter{secnumdepth}{3}
\renewcommand{\thesubsubsection}{S\arabic{subsubsection}}
\subsubsection{Acousto-electric effect for the higher-order mode}
\label{sec:8}
\begin{figure}[!b] 
	\centering
	\includegraphics[width=0.8\textwidth]{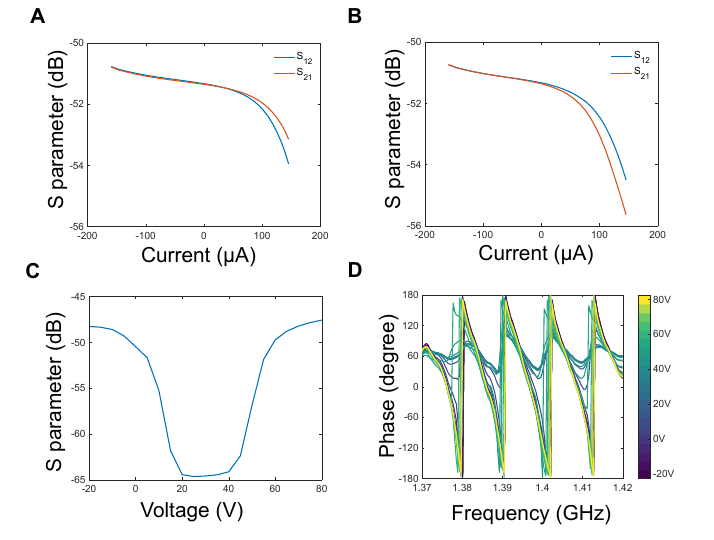} 
	\caption{\textbf{The acousto-electric effect for the higher-order mode.}
		(A) Same as Figure~\ref{fig:S11}(A), but for the higher-order mode. (B) Same as Figure~\ref{fig:S11}(B), but for the higher-order mode. (C) Same as Figure~\ref{fig:3}(B), but for the higher-order mode. (D) Same as Figure~\ref{fig:3}(G), but for the higher-order mode.
		}
	\label{fig:S5} 
\end{figure}
The higher-order mode, which is characterized by oscillatory behavior between the top surface and the buffer-oxide/bottom-silicon interface exhibits the acousto-electric effect similar to that of the Sezawa mode. Fig.~\ref{fig:S5}(A) and Fig.~\ref{fig:S5}(B) demonstrate the nonreciprocal response under an applied drift current. Fig.~\ref{fig:S5}(C) and Fig.~\ref{fig:S5}(D) show the carrier-concentration–induced acousto-electric damping beneath the two IDT regions, along with the corresponding horizontal shift in the phase spectrum, indicating a change in the wave average velocity along the whole delay line.
\setcounter{secnumdepth}{3}
\renewcommand{\thesubsubsection}{S\arabic{subsubsection}}
\subsubsection{Carrier concentration induced acousto-electric effect for different wavelength}
\label{sec:9}
We investigate mode-selective acousto-electric damping to rule out the possibility of electrical-port mismatch, which could reduce the total energy injected into the SAW delay line under a DC bias. In the measurements of these new devices, so the acoustic frequency varies over a modest range (from $1.2~\mathrm{GHz}$ to $1.55 ~\mathrm{GHz}$), the spatial growth rate roughly satisfies
\begin{equation}
    \alpha \propto kL,
    \label{eq:S157}
\end{equation}
which indicates similar acoustic responses when the IDT wavelength is sweeping from $3.8~\mathrm{\mu m}$ to $2.7\mathrm{\mu m}$ ($3.8~\mathrm{\mu m }$ results are shown in the main text) while the number of teeth remain unchanged. Fig.~\ref{fig:S6}(A), Fig.~\ref{fig:S6}(B), Fig.~\ref{fig:S6}(C) show results of $3.6~\mathrm{\mu m}$ that are consistent with those in the main text. The remaining panels also exhibit similar trends. At smaller wavelength, the excitation of acoustic modes becomes weaker and the non-time gated phase signatures are buried inside the electromagnetic feedthrough.
\begin{figure}[!b] 
	\centering
	\includegraphics[width=1\textwidth]{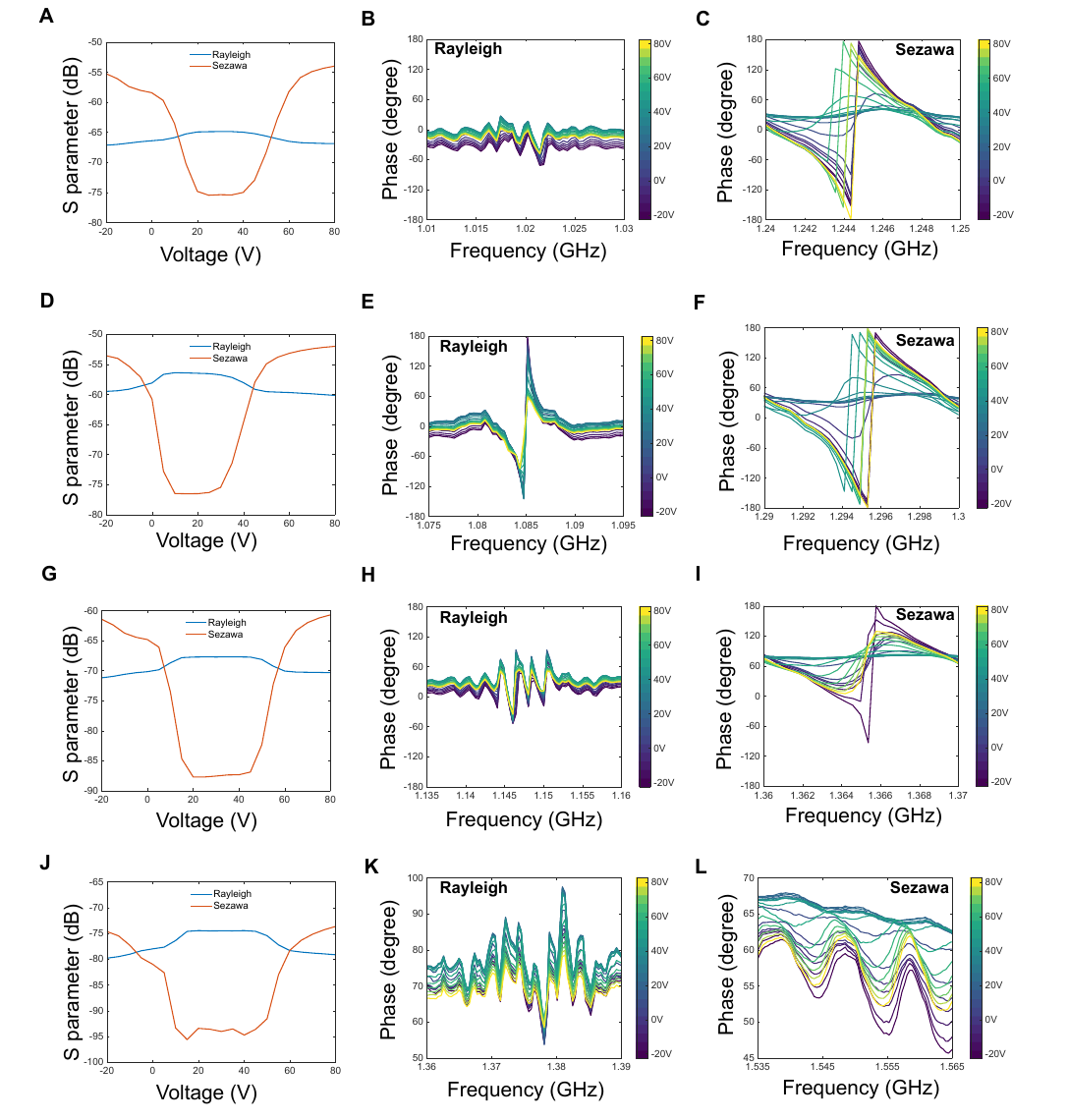} 
	\caption{\textbf{Carrier-induced acousto-electric tuning across IDT wavelengths.}
		For each IDT wavelength $\lambda$, the left/middle/right panels show the same quantities as Figure~\ref{fig:3}B, Figure~\ref{fig:3}G, and Figure~\ref{fig:3}H, respectively.
        (A--C) $\lambda=3.6~\mathrm{\mu m}$.
        (D--F) $\lambda=3.4~\mathrm{\mu m}$.
        (G--I) $\lambda=3.2~\mathrm{\mu m}$.
        (J--L) $\lambda=2.7~\mathrm{\mu m}$.
		}
	\label{fig:S6} 
\end{figure}
\setcounter{secnumdepth}{3}
\renewcommand{\thesubsubsection}{S\arabic{subsubsection}}
\subsubsection{Cross-validation of RF and optical suppression}
\label{sec:10}
Fig.~\ref{fig:S7} compares how the RF transmission and the optical Stokes power change as the bias voltage is stepped. To compare them directly on the same axis, the RF change is plotted with a half scaling in $\mathrm{dB}$. The overlap between two curves indicates the two-transducer RF delay-line experiences an additional AE damping from the receiving IDT.
\begin{figure}[!b] 
	\centering
	\includegraphics[width=1\textwidth]{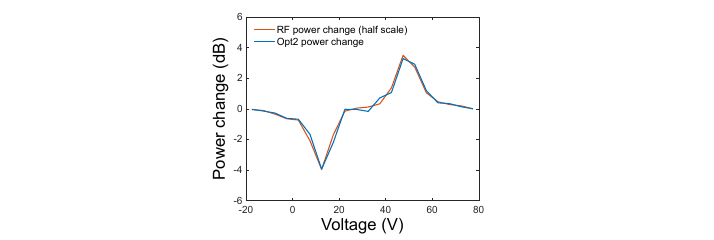} 
	\caption{\textbf{Power change under applied bias.} The data is the same with Figure~\ref{fig:3}B and Figure~\ref{fig:3}C. The y-axis shows the change in power between successive voltage steps, plotted at the midpoint voltage between each pair of adjacent bias points. The derivative from the RF curve is scaled by a half in dB to align with the Opt2 derivative curve for direct comparison.
	}
	\label{fig:S7} 
\end{figure}
\setcounter{secnumdepth}{3}
\renewcommand{\thesubsubsection}{S\arabic{subsubsection}}
\subsubsection{Gate-controlled carrier modulation}
\label{sec:11}
Since the acousto-electric modulation of phonons depends on carrier concentration, the gate voltage that applies a vertical electric field can tune the carrier density. For the same device studied in the main text, the maximum Sezawa damping occurs at $V_g=-40~\mathrm{V}$ in Fig.~\ref{fig:S8}. This minimum transmission is comparable to the case where a lateral electric field is applied using on-chip electrodes.
\begin{figure}[!b] 
	\centering
	\includegraphics[width=1\textwidth]{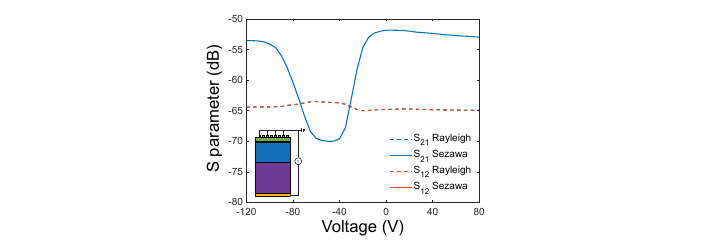} 
	\caption{\textbf{RF S-parameter measurements of the acousto-electric effect induced by gate-controlled carrier concentration.}
		  A gold layer (shown in yellow) is deposited on top of the device stage to enable a vertical electric field to the IDT ground reference.
		}
	\label{fig:S8} 
\end{figure}
\setcounter{secnumdepth}{3}
\renewcommand{\thesubsubsection}{S\arabic{subsubsection}}
\subsubsection{Reconstructed phase spectrum}
\label{sec:12}
To reconstruct the phase spectrum using Equation~\ref{eq:2}, the amplitude of the EM feedthrough is roughly estimated at $-60~\mathrm{dB}$ based on Fig.~\ref{fig:S4}(C). The center frequency $\omega_0$ is extracted from Figure~\ref{fig:2}C and the effective number of IDT fingers $N_p$ is estimated based on the envelop of the phase spectrum. The voltage-dependent acoustic amplitude is extracted from Figure~\ref{fig:3}B and the background phase shift $\phi_0$ can be extracted out on the neighboring frequency of the resonant Sezawa mode. The effective time delay induced by AE, the unperturbed time delay and the accurate amplitude of the EM feedthrough are defined when the reconstructed spectrum matches the peak feature from $1.2~\mathrm{GHz}$ to $1.21~\mathrm{GHz}$, resulting in $-59.39~\mathrm{dBm}$ EM feedthrough power and $0.18 ~\mathrm{ns}$ maximum time delay induced by AE out of total $114.155~\mathrm{ns}$ unperturbed signal delay. Moreover, $N_p$ is $43$ and the center frequency of the Sezawa mode is $1.2055~\mathrm{GHz}$ 
\setcounter{secnumdepth}{3}
\renewcommand{\thesubsubsection}{S\arabic{subsubsection}}
\subsubsection{Amplitude-phase modulation}
\label{sec:13}
\begin{figure}[!b] 
	\centering
	\includegraphics[width=1\textwidth]{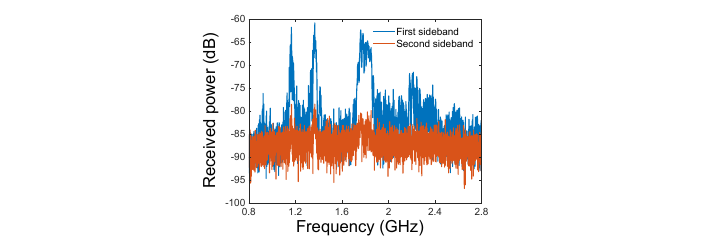} 
	\caption{\textbf{Amplitude-phase modulation by optical readout.}
		  The setup is the same with Figure~\ref{fig:4}A, but for $\mathrm{V_{pp}}=5~\mathrm{V}$ at $100~\mathrm{kHz}$ modulation frequency.
		}
	\label{fig:S9} 
\end{figure}
Since the acousto-electric effect tunes both amplitude and phase, we investigate the small-signal modulation model. The modulated signal has the form
\begin{equation}
    s(t)=A(1+\varepsilon\cos\omega t)\cos(\Omega t +\phi_0 +\beta\cos\omega t),
    \label{eq:S158}
\end{equation}
where $\Omega$ is the angular frequency of the acoustic wave; $\omega$ is the modulation frequency; $\phi_0$ is the initial phase set by the working voltage bias $V_0$; $\varepsilon$ and $\beta$ are the amplitude modulation depth (dimensionless) and the phase modulation index (in radians); $A$ is the amplitude of the acoustic wave set by the working voltage bias. Use Jacobi-Anger expansion
\begin{equation}
    \cos(\Omega t+\phi_0+\beta\cos\omega t)=\Re\left\{e^{i(\Omega t+\phi_0)}\sum_{n=-{\infty}}^{\infty}i^nJ_n(\beta)e^{in\omega t}\right\},
    \label{eq:S159}
\end{equation}
the coefficient is
\begin{equation}
    C_m=Ae^{i\phi_0}(i^m J_m(\beta)+\frac{\varepsilon}{2}(i^{m-1}J_{m-1}(\beta)+i^{m+1}J_{m+1}(\beta))),
    \label{eq:S160}
\end{equation}
with
\begin{equation}
    s(t)=\Re\left\{\sum_{m=-\infty}^{\infty}C_me^{i(\Omega+m\omega)t}\right\}.
    \label{eq:S161}
\end{equation}
Since
\begin{equation}
    J_m'(\beta)=\frac{1}{2}(J_{m-1}-J_{m+1}),
    \label{eq:S162}
\end{equation}
the coefficient becomes
\begin{equation}
    C_m=Ae^{i\phi_0}i^m(J_m(\beta)-i\varepsilon J_m'(\beta)),
    \label{eq:S163}
\end{equation}
with
\begin{equation}
    |C_m|=A\sqrt{(J_m(\beta))^2+(\varepsilon J_m'(\beta))^2}.
    \label{eq:S164}
\end{equation}
At small amplitude modulation depth and phase modulation index
\begin{equation}
    |C_1|\approx\frac{A}{2}\sqrt{\beta^2+\varepsilon^2}.
    \label{eq:S165}
\end{equation}
The modulation spectrum should resemble the pure phase modulation, but with the modification of the amplitude and the phase angle. In Fig.~\ref{fig:S9}, when the modulation voltage increases to $\mathrm{V_{pp}=5~V}$, a second sideband appears in the spectrum, indicating mixed amplitude–phase modulation caused by acoustoelectric-induced acoustic damping and the accompanying change in acoustic velocity.
\setcounter{secnumdepth}{3}
\renewcommand{\thesubsubsection}{S\arabic{subsubsection}}
\subsubsection{Ohmic contact}
\label{sec:14}
Ohmic contact reduces contact resistance, which facilitates current flow and decreases the bias voltage required to tune the carrier concentration. We investigate three doping configurations on the AlN-SOI platform. As shown in Fig.~\ref{fig:S10}, only $\mathrm{n+nn+}$ doping yields an ohmic Al/Si contact. In this case, the heavily doped silicon under Al electrodes strongly bends the band structure and reduces the barrier width, making tunneling significant that forms the Ohmic contact. Without the heavy doping region, the Schottky barriers are formed due to the mismatch work functions between Si and Al. However, for the $\mathrm{p+pp+}$ doping, depletion regions are still observed. We attribute this to not sufficient heavy doping concentration. As is indicated by Paul \textit{et al.} \cite{paul2020single}, a much heavier p doping concentration is required to form the Ohmic contact.
\begin{figure}[!b] 
	\centering
	\includegraphics[width=1\textwidth]{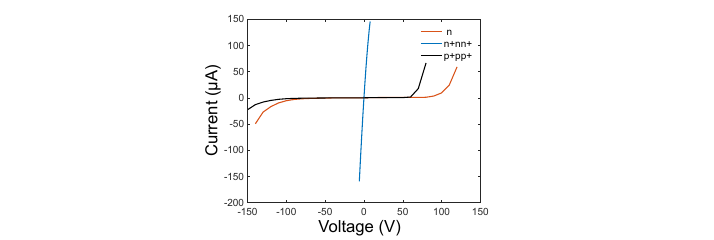} 
	\caption{\textbf{IV curves for different types of doping.}
		  n: $1\times 10^{16}~\mathrm{cm^{-3}}$ n doping along the acoustic delay line. n+nn+:$8\times 10^{19}~\mathrm{cm^{-3}}$ heavy n doping under aluminum electrodes and $1\times 10^{16}~\mathrm{cm^{-3}}$ n doping along the acoustic delay line. p+pp+: $8\times 10^{19}~\mathrm{cm^{-3}}$ heavy p doping under aluminum electrodes and $1\times 10^{16}~\mathrm{cm^{-3}}$ p doping along the acoustic delay line.
		}
	\label{fig:S10} 
\end{figure}
\setcounter{secnumdepth}{3}
\renewcommand{\thesubsubsection}{S\arabic{subsubsection}}
\subsubsection{Selective current-induced AE effect}
\label{sec:15}
\begin{figure}[!b] 
	\centering
	\includegraphics[width=0.8\textwidth]{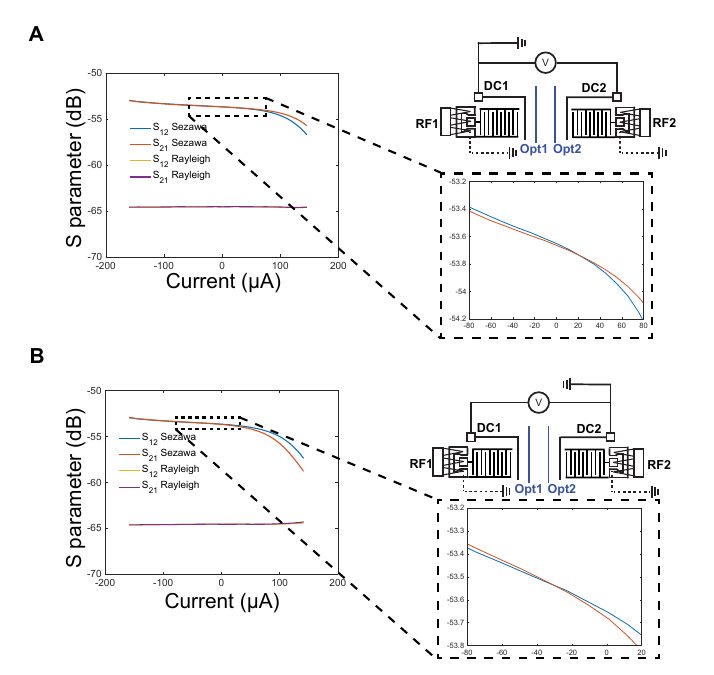} 

	\caption{\textbf{Current–induced nonreciprocity via the AE effect}
		(\textbf{A}) Time-gated RF S-parameter measurements over a drift-current sweep with DC1 grounded. The labeling scheme is consistent with that used in Figure \ref{fig:2}B, the RF port assignments (RF1 and RF2) are interchanged.  (\textbf{B}) Time-gated RF S-parameter measurements over a drift-current sweep with DC2 grounded. The labeling scheme is consistent with that used in Figure \ref{fig:2}B, except that the RF port assignments (RF1 and RF2) are interchanged. }
	\label{fig:S11} 
\end{figure}
A defining feature of the AE interaction is its nonreciprocal nature. According to Equation~\ref{eq:1} of the main text, the carrier drift can either supply energy to, or extract energy from, the acoustic wave depending on the relative velocity between the carriers and the acoustic wave. If the carrier velocity is larger than the acoustic velocity while they co-propagate, the acoustic wave is effectively dragged by the carriers and exhibits gain (a positive gain coefficient). If the carrier velocity is smaller, or directed opposite to the acoustic propagation, the interaction instead introduces extra loss. Therefore, under an applied current along the delay line, the transmission for forward and backward acoustic propagation relative to the carrier velocity is different, leading to nonreciprocity that can be measured in S-parameters. In Figure~\ref{fig:S11}A, a DC bias applied between DC1 and DC2 drives a proportional current through the n+nn+ formed ohmic contact \cite{paul2020single} in between, causing $\mathrm{S_{12}}\neq \mathrm{S_{21}}$ for the Sezawa mode (and for higher-order modes, though they are not shown here). When the current direction is reversed by grounding DC2 instead of DC1 (Figure~\ref{fig:S11}B), $\mathrm{S_{12}}$ and $\mathrm{S_{21}}$ flip their behaviors, where $\mathrm{S_{12}}$ receives AE gain and $\mathrm{S_{21}}$ experiences additional AE attenuation. A maximum nonreciprocal transmission ratio of $1.4~\mathrm{dB}$ ($4.7~\mathrm{dB/mm}$) is observed for the Sezawa mode in the delay line, accompanied by a significant background tuning of S-parameters due to carrier-concentration change under the IDT regions.


\clearpage 


\renewcommand\refname{References}

\bibliographystyle{sciencemag}
\end{document}